%% file: main.tex
\documentclass[letterpaper,twocolumn,10pt]{article}
\usepackage{usenix}
\input{header}
\usepackage{graphicx}
\usepackage{subcaption}
\usepackage{algorithm}
\usepackage{adjustbox}
\usepackage{algorithmicx}
\usepackage{algpseudocode}
\usepackage{multirow}
\usepackage[table,xcdraw]{xcolor}
\hypersetup{
  colorlinks,
  linkcolor={red!70!black},
  urlcolor={blue!80!black}
}

\usepackage{pgfplots}
\usepackage{subcaption}
\usetikzlibrary{patterns}
\pgfplotsset{compat=1.18}

\algrenewcommand\algorithmiccomment[1]{\hfill\textcolor{blue!70!black}{\(\triangleright\) #1}}
\newcommand{\spab}{{\textsf{SpAB}}\xspace}
\newcommand{\abc}{{\textsc{Maui}}\xspace}
\newcommand{\rulesep}{\unskip\ \vrule\ }
\begin{document}
\date{}

\title{\Large \bf \abc: Reconstructing Private Client Data in Federated Transfer Learning}

\author{
{\rm Ahaan Dabholkar}\\
Purdue University
\and
{\rm Atul Sharma}\\
Purdue University
\and
{\rm Z. Berkay Celik}\\
Purdue University
\and
{\rm Saurabh Bagchi}\\
Purdue University\\
\and{\rm\normalsize\text{
\{adabholk, sharm438, zcelik, sbagchi\}@purdue.edu
}}
} 

\maketitle

\begin{abstract}
Recent works in federated learning~(FL) have shown the utility of leveraging transfer learning
for balancing the benefits of FL and centralized learning. In this setting, federated training
happens after a stable point has been reached through conventional training.
Global model weights are first centrally pretrained by the server on a public dataset following
which only the last few linear layers~(the classification head) of the model are finetuned 
across clients.
In this scenario, existing data reconstruction attacks~(DRAs) in FL show two key weaknesses.
First, strongly input-correlated gradient information from the initial model layers is never shared,
significantly degrading reconstruction accuracy.
Second, DRAs in which the server makes highly specific, handcrafted manipulations to the model 
structure or parameters~(for e.g., layers with all zero weights, identity mappings and rows with
identical weight patterns) are easily detectable by an active client.

Improving on these, we propose \abc, a \textit{stealthy} DRA that does not require any overt 
manipulations to the model architecture or weights, and relies solely on the gradients of the 
classification head.
\abc first extracts ``robust'' feature representations of the input batch from the gradients of
the classification head and subsequently inverts these representations to the original inputs. 
We report highly accurate reconstructions on the CIFAR10 and ImageNet datasets on a variety of
model architectures including convolution networks~(CNN, VGG11), ResNets~(18, 50), ShuffleNet-V2
and Vision Transformer~(ViT B-32), regardless of the batch size. 
\abc significantly outperforms prior DRAs in reconstruction quality, 
achieving 40–120\% higher PSNR scores.
\end{abstract}

\input{sections/1-introduction}
\input{sections/2-background}
\input{sections/3-motivation}
\input{sections/4-threat_model}
\input{sections/5-methodology}
\input{sections/6-implementation}
\input{sections/7-evaluation}
\input{sections/8-discussion}
\input{sections/9-conclusion}

\cleardoublepage
\appendix
\cleardoublepage
\bibliographystyle{plain}
\bibliography{bibfile}

\appendix
\input{appendix/appendix}

\end{document}

%% file: header.tex
\usepackage{tikz}
\usetikzlibrary{positioning, arrows.meta, fit, shapes.geometric, backgrounds}
\usepackage{amsmath,amsfonts}
\usepackage{bbm}
\usepackage{todonotes} 
\usepackage{tikz}
\usepackage{xspace}
\usepackage{xurl}
\usepackage{enumitem}
\usepackage{fixltx2e}
\usepackage{array}
\usepackage{hyperref}

\DeclareRobustCommand{\ballnumber}[1]{\tikz[baseline=(myanchor.base)]
\node[circle,fill=.,inner sep=1pt] (myanchor) {\color{-.}\bfseries\footnotesize #1};}

\newcommand{\highlightbox}[1]{\todo[inline,color=white]{#1}}

\DeclareMathOperator*{\argmin}{arg\,min}

\newcommand{\ie}{{i.e.,}\xspace}
\newcommand{\etal}{{\em et al.}\xspace}

\usepackage{microtype}
\usepackage{comment}
\newcommand{\rev}[1]{#1}

%% file: sections/1-introduction.tex
\section{Introduction}
Federated learning~(FL) is a decentralized learning protocol that enables collaborative training of a model
across a network of clients, while keeping local data on-device.
Instead of directly sharing private training data with a central server, clients train locally 
to generate gradient updates, which they periodically share with the server. 
The server aggregates these updates and uses them to advance training of the global model.
Studies~\cite{arjevani2015communication, liconvergence, woodworth2020local} have shown that 
compared to centralized training, FL is slower to converge and often results in inferior models
due to the non-iid data distributions across clients.

Federated transfer learning (FTL) has recently emerged as a promising solution to mitigate these
issues~\cite{gao2019privacy, kevin2021federated, liu2020secure}. While vanilla FL starts training
with random model parameters, FTL initializes the model with parameters learned centrally on a public dataset.
This pretrained model is then finetuned through a conventional federated training process~(FedSGD) 
on the clients' private datasets. FTL has been shown to be more resilient to the data and device 
heterogeneity present across clients and leads to higher training accuracies with faster convergence 
than vanilla federated learning~\cite{chenimportance,nguyenbegin}.

However, the gradient updates shared by clients can be exploited by an attacker-controlled server 
to mount \emph{data reconstruction attacks}~(DRAs). DRAs aim to recover representations of a 
client's private training batch during federated training.
Early DRAs~\cite{zhu2019deep,zhao2020idlg,geiping2020inverting,yin2021see} relied on an optimization-based 
approach called gradient matching to reconstruct a client's input batch. In this, the server 
iteratively optimizes a candidate batch to minimize the difference between the received and generated gradients. 
Other methods~\cite{geiping2020inverting,yin2021see} incorporate domain specific priors into the optimization, 
leading to more accurate reconstructions. Such DRAs operate under an honest-but-curious server 
threat model in which the server is limited to performing offline analyses on the received gradient 
updates. These DRAs are largely unsuccessful when the client's batch size or the complexity of the model grows.
{\em While it may appear that DRAs against vanilla FL would be equally applicable against FTL, that is not the case.}
FTL typically involves clients sharing only a fraction of the model's gradients, degrading the 
accuracy of gradient matching.

To overcome the weaknesses of optimization-based DRAs on large batch sizes or complex models, 
recent DRAs~\cite{fowl2022robbing,pan2022exploring,boenisch2023curious,zhao2024loki,shi2025scale} 
assume a malicious server to mount more effective attacks. In this setting, the server strategically 
handcrafts the model parameters sent in any round of federated training or, in extreme cases, 
modifies the model's structure to simplify data reconstruction. For example, a recent work uses 
an identity distribution for convolution kernel weights, which contains a single non-zero element 
in the weight vector~\cite{boenisch2023curious}. Another line of work modifies the structure of 
the model by adding linear layers with rows of identical weights to the \textit{front} of the 
feature extractor~\cite{fowl2022robbing, zhao2024loki}. 
We identify four such \emph{privacy-leaking primitives} that are used singly or in combination 
by all prior works that fall in the malicious server category. 
We show that these privacy-leaking primitives exhibit parameter distributions fundamentally 
distinct from those observed during normal model training. Therefore, an active client can 
detect a potential DRA through simple anomaly detection on the received model weights or model 
structure inspection.

We propose a new DRA, \abc, specifically targeted to the  federated transfer learning setting 
and which does not rely on any privacy-leaking primitives.
This setting better reflects real-world attack scenarios where overt manipulations of the model 
are likely to be noticed and countered by clients.
Inspired by~\cite{pan2022exploring}, we propose a data reconstruction method, that uses 
intermediate representations~(IRs) of a client's batch produced by the model's feature extractor, 
for reconstruction.
\abc is based on two key ideas which together result in large scale data recovery during FTL. 
First, we impose a strong prior on the IRs generated by the feature extractor using 
``adversarially robust'' parameters, rather than strong priors on the reconstructed samples as 
done in all prior works. 
Second, we learn parameters for the model's classification head that enforces a property called 
{\em column sparsity}, which allows for faithful recovery of these IRs from their gradients.
This is achieved by optimizing a novel loss function that enhances the conditions required for 
successful linear layer leakage.
On receiving the gradient update,  IRs of all samples of the batch are extracted and inverted in 
parallel using an IR-matching optimization loop, thereby making the reconstruction independent of 
the client's batch size. Here, the imposition of the robust prior on the IRs dramatically improves 
the quality of the final reconstruction, especially for large datasets like ImageNet and on complex 
model architectures like vision transformers~(ViT B-32) and ResNets~(18,50). 

We conduct extensive experiments across diverse datasets (CIFAR10, ImageNet, MedMNIST, CelebA) 
and model architectures spanning shallow networks like - CNN, ResNet18, ShuffleNet-V2 and deep 
networks like ResNet50, VGG11 and ViT B-32. We achieve significantly better reconstruction quality, 
as measured by PSNR, SSIM, and LPIPS scores, compared to state-of-the-art baselines~(those that do 
not use any of the privacy-leaking primitives), namely DLG~\cite{zhu2019deep}, Inverting Gradients~\cite{geiping2020inverting}, 
and Neuron Exclusivity~\cite{pan2022exploring}. Moreover, this reconstruction quality is preserved even for large batch sizes. 

\vspace*{0.5em}
\noindent\textbf{Summary of Contributions.}
To summarize, we make the following contributions in this paper.
\begin{itemize}[leftmargin=*]
  \item We identify four privacy-leaking primitives used by prior DRAs operating within a malicious 
  server threat model. We show these primitives present anomalous parameter distributions and are 
  easily detectable by an active client. 
  \item We introduce a novel IR-matching based DRA, \abc, uniquely tailored to FTL which does not 
  rely on any privacy leaking primitives and uses robust priors for reconstruction.
  \item We propose a novel training method called \spab-training for the model's classification head 
  which significantly increases the probability of recovering IRs for a large fraction of the client's 
  batch from the received gradient update. 
\end{itemize}

%% file: sections/2-background.tex
\section{Background}
\label{sec:background}
\input{tables/intro-table.tex}

\subsection{Federated Transfer Learning}
\label{sec:federated-learning}
Federated learning~(FL)~\cite{mcmahan2017communication} is a decentralized learning protocol 
that enables a network of clients to learn a global model while keeping their datasets private. 
FL presents certain drawbacks when compared to centralized training. The data and device
heterogeneity present across clients typically reduces the performance of FL training. The exchange
of parameters and gradient updates further incurs a communication cost. 
Transfer learning has recently emerged as a potential solution to both these issues. 
Instead of initializing the global model with random parameters at the beginning of federated 
training, the server uses weights pretrained on a public dataset as the initial parameters. 
Communication overhead is minimized by only finetuning the final layers of the model in a 
federated fashion. This approach of centralized pretraining + federated finetuning has been 
shown to significantly reduce convergence time and improve the accuracy of the final 
model~\cite{nguyenbegin,legate2023guiding}.

Consider a network of $n$ participating clients with private datasets $\{D^\textsf{priv}_i\}_{i=1}^n$ 
that wish to train a model $h$ using FTL. Let $h$ be composed of the feature extractor $h_F(.;\theta_F)$
and the classification head $h_C(.; \theta_C)$ as $h(.;\theta) := h_C(h_F(.;\theta_F);\theta_C)$. 
First, in the pretraining phase, the server initializes $h$ by training on a public dataset $D^\textsf{pub}$.
Alternatively, the server downloads a pretrained model having feature extractor $h_F$ and replaces 
the existing classification head with $h_C$. This is often done to account for the differences 
between $D^\textsf{pub}$ and the FL task~(for eg, replacing the $1000$ label classification head 
of a public ImageNet model with a fewer label head to match the FL task). 
The trained parameters of the feature extractor, $\theta^\textsf{pub}_F$ are then broadcast to the 
clients once and frozen. Next, the federated finetuning of the classification head 
proceeds iteratively over a number of rounds using FedSGD. In round $t$, the server broadcasts 
parameters $\theta_C^{(t)}$ to the participating clients. The $i^{th}$ client uses a batch of private 
samples and labels $(\mathbf X, \mathbf C) \sim D^{\textsf{priv}}_i$ of size $B_i$ to locally compute 
the gradient update 
$
  \nabla_{\theta_C} L\big(h_C(h_F(\mathbf X;\theta^\textsf{pub}_F);\theta^{(t)}_C), \mathbf C\big)
$
for the scalar loss $L$. Each client then uploads their update to the server which aggregates them
using a weighted mean and computes $\theta_C^{(t+1)}$.

\subsection{Data Reconstruction Attacks in FL}
\label{sec:data-reconstruction-attacks}
Data reconstruction attacks~(DRAs) in FL are server-mounted attacks that breach client privacy 
by reconstructing private training samples during federated training. The server is assumed 
to be controlled by an attacker and uses the gradient updates received from clients to mount the
attack. \rev{In the context of this work, prior DRAs from FL can be mounted during the federated 
finetuning phase of FTL.}
DRAs are classified into two categories based on their method of reconstruction.

\subsubsection{Optimization-based DRAs}
\label{sec:optimization_attacks}
These DRAs~(\cite{zhu2019deep,zhao2020idlg,geiping2020inverting,yin2021see}) reconstruct a client's
input batch by solving a ``gradient matching'' objective using standard optimization techniques 
like gradient descent. If we consider that gradient update $g$ is received at the end of round $t$,
the gradient-matching objective can be expressed as follows
{\small
\begin{align}
\label{eqn:gradient-matching}
  \mathbf X^*, \mathbf C^* = \argmin_{\mathbf X,\mathbf C} \Big[ 
    \mathcal D\big(\nabla_\theta L\big(h(\mathbf X;\theta^{(t)}), \mathbf C\big), g \big) 
    + \mathcal R(\mathbf X) 
  \Big]
\end{align}
}%
The server tries to learn the optimal sample-label batch 
$\mathbf X^*, \mathbf C^*$ which minimizes the $\mathcal D$-distance between the computed gradient
and $g$. However, this reconstruction method fails to generate accurate reconstructions when the
client batch size is large or when the datasets and models are complex. Improvements made 
by~\cite{zhao2020idlg, yin2021see, geiping2020inverting} simplify gradient-matching by first 
recovering $\mathbf C^*$. More recent DRAs use regularizers~($\mathcal R$) to add prior knowledge
to the optimization process further enhancing reconstruction accuracy.
Examples of popular regularizers are TV norm prior~\cite{geiping2020inverting}, batch 
normalization~(BN), DeepInversion~(DI) and group consistency~(GC) prior~\cite{yin2021see} to 
improve the reconstruction quality for natural image datasets.

\subsubsection{Analytic DRAs}
\label{sec:analytic_attacks}
Optimization-based DRAs are limited by the intractability of gradient-matching which heavily depends
on the complexity of the network, the size of the input space and even the contents of the batch 
itself.
For instance, the label distribution of the batch has been shown to significantly affect the 
reconstruction quality~\cite{zhao2024leak}. 
Improving on these, a more recent line of work proposes an analytic approach to reconstruction 
by establishing an invertible relation between the client's batch and the gradients of specific 
model layers.
Specifically, \cite{phong2017privacy,zhu2020rgap} show that the weight and bias gradients of a 
linear layer can be used to \textit{exactly} recover the input for single sample batches. 
Recent works~\cite{fowl2022robbing,boenisch2023curious,zhao2024loki} extend this mechanism of 
\textit{linear layer leakage} to enable the exact recovery of multi-sample batches. 
Despite their effectiveness, we show that analytic DRAs present certain disadvantages which limit
their feasibility during FTL.

%% file: tables/intro-table.tex
\begin{table*}[t!]
\caption{Summary of DRAs in FL. \colorbox[HTML]{FFCCC9}{\phantom{ah}} shows the usage of the privacy leaking primitives or the dependencies of the respective DRAs. Recent analytic attacks like Robbing the Fed~(RtF)~\cite{fowl2022robbing}, When the Curious Abandon Honesty~(CAH)~\cite{boenisch2023curious}, Fishing for User Data in Large Batch Federated Learning~(Fishing)~\cite{wen2022fishing}, Loki~\cite{zhao2024loki} and Scale-MIA~\cite{shi2025scale} all rely on using atleast one privacy primitive to be effective and have some dependencies. For the definitions of these primitives, we refer the reader to \autoref{sec:privacy-leaking-primitives}.}
\centering
\resizebox{0.95\textwidth}{!}{%
\renewcommand{\arraystretch}{1.1}
\begin{tabular}{|c|c|c|c|cccc|cc|}
\hline
 &  &  &  & \multicolumn{4}{c|}{\textbf{Privacy Leaking Primitives}} & \multicolumn{2}{c|}{\textbf{Attack Dependencies}} \\ \cline{5-10} 
\multirow{-2}{*}{\textbf{Attack}} & \multirow{-2}{*}{\textbf{\begin{tabular}[c]{@{}c@{}}Attack \\ Type\end{tabular}}} & \multirow{-2}{*}{\textbf{\begin{tabular}[c]{@{}c@{}}Attacker \\ Capability\end{tabular}}} & \multirow{-2}{*}{\textbf{\begin{tabular}[c]{@{}c@{}}Prior\\ Introduced\end{tabular}}} & \multicolumn{1}{c|}{\textbf{\begin{tabular}[c]{@{}c@{}}Structural\\ Modification\end{tabular}}} & \multicolumn{1}{c|}{\textbf{\begin{tabular}[c]{@{}c@{}}RtF \\ Primitive\end{tabular}}} & \multicolumn{1}{c|}{\textbf{\begin{tabular}[c]{@{}c@{}}Identity Mapping\\ Kernels\end{tabular}}} & \textbf{\begin{tabular}[c]{@{}c@{}}All Zeros \\ Kernels\end{tabular}} & \multicolumn{1}{c|}{\textbf{\begin{tabular}[c]{@{}c@{}}Model \\Structure\end{tabular}}} & \textbf{\begin{tabular}[c]{@{}c@{}}Label\\Distribution\end{tabular}} \\ \hline
DLG~\cite{zhu2019deep}/iDLG~\cite{zhao2020idlg} & Optimization & Honest but Curious & - & \multicolumn{1}{c|}{\cellcolor[HTML]{FFFFFF}} & \multicolumn{1}{c|}{\cellcolor[HTML]{FFFFFF}} & \multicolumn{1}{c|}{\cellcolor[HTML]{FFFFFF}} & \cellcolor[HTML]{FFFFFF} & \multicolumn{1}{c|}{\cellcolor[HTML]{FFFFFF}} & \cellcolor[HTML]{FFCCC9} \\ \hline
IG~\cite{geiping2020inverting} & Optimization & Honest but Curious & TV & \multicolumn{1}{c|}{\cellcolor[HTML]{FFFFFF}} & \multicolumn{1}{c|}{\cellcolor[HTML]{FFFFFF}} & \multicolumn{1}{c|}{\cellcolor[HTML]{FFFFFF}} & \cellcolor[HTML]{FFFFFF} & \multicolumn{1}{c|}{\cellcolor[HTML]{FFFFFF}} & \cellcolor[HTML]{FFCCC9} \\ \hline
GradInversion~\cite{yin2021see} & Optimization & Honest but Curious & BN, DI, GC & \multicolumn{1}{c|}{\cellcolor[HTML]{FFFFFF}} & \multicolumn{1}{c|}{\cellcolor[HTML]{FFFFFF}} & \multicolumn{1}{c|}{\cellcolor[HTML]{FFFFFF}} & \cellcolor[HTML]{FFFFFF} & \multicolumn{1}{c|}{\cellcolor[HTML]{FFFFFF}} & \cellcolor[HTML]{FFCCC9} \\ \hline
RtF~\cite{fowl2022robbing} & Analytic & Malicious & - & \multicolumn{1}{c|}{\cellcolor[HTML]{FFCCC9}} & \multicolumn{1}{c|}{\cellcolor[HTML]{FFCCC9}} & \multicolumn{1}{c|}{\cellcolor[HTML]{FFCCC9}} & \cellcolor[HTML]{FFFFFF} & \multicolumn{1}{c|}{\cellcolor[HTML]{FFCCC9}} & \cellcolor[HTML]{FFFFFF} \\ \hline
CAH~\cite{boenisch2023curious} & Analytic & Malicious & - & \multicolumn{1}{c|}{\cellcolor[HTML]{FFFFFF}} & \multicolumn{1}{c|}{\cellcolor[HTML]{FFFFFF}} & \multicolumn{1}{c|}{\cellcolor[HTML]{FFCCC9}} & \cellcolor[HTML]{FFCCC9} & \multicolumn{1}{c|}{\cellcolor[HTML]{FFCCC9}} & \cellcolor[HTML]{FFFFFF} \\ \hline
Loki~\cite{zhao2024loki} & Analytic & Malicious & - & \multicolumn{1}{c|}{\cellcolor[HTML]{FFCCC9}} & \multicolumn{1}{c|}{\cellcolor[HTML]{FFCCC9}} & \multicolumn{1}{c|}{\cellcolor[HTML]{FFCCC9}} & \cellcolor[HTML]{FFCCC9} & \multicolumn{1}{c|}{\cellcolor[HTML]{FFCCC9}} & \cellcolor[HTML]{FFFFFF} \\ \hline
Scale-MIA~\cite{shi2025scale} & Hybrid & Malicious &  & \multicolumn{1}{c|}{\cellcolor[HTML]{FFFFFF}} & \multicolumn{1}{c|}{\cellcolor[HTML]{FFCCC9}} & \multicolumn{1}{c|}{\cellcolor[HTML]{FFFFFF}} & \cellcolor[HTML]{FFFFFF} & \multicolumn{1}{c|}{\cellcolor[HTML]{FFFFFF}} & \cellcolor[HTML]{FFFFFF} \\ \hline
NEX~\cite{pan2022exploring} & Hybrid & (Weak)~Malicious & DIP & \multicolumn{1}{c|}{\cellcolor[HTML]{FFFFFF}} & \multicolumn{1}{c|}{\cellcolor[HTML]{FFFFFF}} & \multicolumn{1}{c|}{\cellcolor[HTML]{FFFFFF}} & \cellcolor[HTML]{FFFFFF} & \multicolumn{1}{c|}{\cellcolor[HTML]{FFFFFF}} & \cellcolor[HTML]{FFFFFF} \\ \hline \hline
\textbf{\abc} & Hybrid & (Weak)~Malicious & Robust & \multicolumn{1}{c|}{\cellcolor[HTML]{FFFFFF}} & \multicolumn{1}{c|}{\cellcolor[HTML]{FFFFFF}} & \multicolumn{1}{c|}{\cellcolor[HTML]{FFFFFF}} & \cellcolor[HTML]{FFFFFF} & \multicolumn{1}{c|}{\cellcolor[HTML]{FFFFFF}} & \cellcolor[HTML]{FFFFFF} \\ \hline
\end{tabular}%
}
\label{tab:intro-table}
\end{table*}

%% file: sections/3-motivation.tex
\section{Motivation -- The Flaws of Existing DRAs}
\label{sec:motivation}
\noindent\textbf{Applicability to FTL.}
To generate accurate reconstructions, DRAs require gradients which are strongly correlated with
the input batch. Gradients of initial model layers are thus more important as they carry a more
direct imprint of the raw inputs than gradients of deeper layers~\cite{fowl2022robbing,zhao2024loki}. 
But in FTL, since clients only share gradients of the last few layers~(typically the classification
head), DRA performance suffers significantly.

\noindent{\textbf{Structural Dependence.}}
Analytic DRAs leverage specific features of a model's architecture to simplify data reconstruction.
While this improves reconstruction accuracy, it limits their applicability to specific model 
structures. For instance, skip connections in a ResNet and specialized parameterizations of convolution 
kernels in a CNN are used by~\cite{boenisch2023curious} to mount a powerful DRA. However, 
mounting~\cite{boenisch2023curious} on structurally different models, like Vision Transformers 
is infeasible. This is in contrast to optimization-based DRAs which are more broadly applicable 
to general model architectures where second-order input gradients can be efficiently computed. 

\noindent\textbf{Detectability.}
Analytic DRAs~\cite{fowl2022robbing,boenisch2023curious,zhao2024loki,shi2025scale,wen2022fishing}
operate under the malicious server threat model, where the server has the ability to send 
arbitrary parameters to a client or make unchecked modifications to the model architecture.
We show that such manipulations present highly anomalous, handcrafted patterns that are unlikely
to occur during genuine model training, thus making them easily detectable to an active client.
We collectively refer to these mechanisms as \textit{``privacy-leaking primitives''}.
We analyze recent works in FL literature and compile a list of privacy-leaking
primitives. According to our analysis, most of the recently proposed DRAs make extensive use of
these primitives, thus limiting their real-world effectiveness.
\autoref{tab:intro-table} summarizes our analysis and we present detailed descriptions
of each primitive in~\autoref{sec:privacy-leaking-primitives}. 

We motivate our work by overcoming these three drawbacks and presenting an architecture 
independent DRA on FTL which does not rely on \textit{any} privacy-leaking primitives.

\subsection{Privacy-Leaking Primitives}
\label{sec:privacy-leaking-primitives}
\noindent\ballnumber{1}~\underline{\textit{Identity-Mapping Kernel.}}
This primitive is a 2D convolution kernel with kernel weights that impose an identity mapping 
between one or more input-output channels. For the $i^{th}$ input channel, the $(i, k/2, k/2)$
element of a single $k \times k$ kernel is set to $1$ and all other elements are set to $0$. 
This primitive is used by DRAs to propagate inputs without distortion through the feature-extractors
of convolution networks in~\cite{fowl2022robbing,boenisch2023curious,zhao2024loki}.

\noindent\ballnumber{2}~\underline{\textit{RtF Primitive.}}
Fowl \etal~\cite{fowl2022robbing} propose a module with two high-dimensional linear layers separated
by a ReLU or ReLU6 activation. The weight and bias parameters of this module sent by the server 
enable a large scale extraction of the input batch from the module's gradients. The first layer
is initialized such that each row of its weight vector is identical. The second layer is initialized
to be column-wise identical~(for details, see~\cite{fowl2022robbing}). This establishes an invertible
relation between the inputs to the module and its gradients, allowing closed-form
input recovery.
This primitive is critical for DRAs~\cite{fowl2022robbing,zhao2024loki,shi2025scale}.  

\noindent\ballnumber{3}~\underline{\textit{All-Zeros Kernel.}}
The all zeros kernel is simply a 2D convolution kernel with all kernel weights set to $0$. 
It is used by~\cite{zhao2024loki} to prevent the mixing of gradients from different clients 
during secure aggregation, thereby scaling the RtF~\cite{fowl2022robbing} DRA to a larger 
number of clients. It is also used to prevent input distortion from residual connections in 
ResNets by~\cite{boenisch2023curious}.

\noindent\ballnumber{4}~\underline{\textit{Structural Modification.}}
By participating in FL training, clients implicitly agree on the architecture of the model that
is being trained. Conventionally, an image classification architecture comprises a feature-extractor
made of convolution layers and an MLP as the classification head. DRAs presented 
in~\cite{fowl2022robbing,zhao2024loki} modify this standard, pre-agreed architecture by introducing
high-dimensional~(input-sized) linear layers at the beginning of the model. We refer to any such
unusual modifications of the model as the structural modification primitive.

\subsection{Detecting Privacy-Leaking Primitives}
\label{sec:anomaly-detector}
\begin{figure}[t!]
\centering
\begin{subfigure}{0.40\linewidth}
  \includegraphics[width=\linewidth,trim={0.em 0em 0.em 0.em},clip]{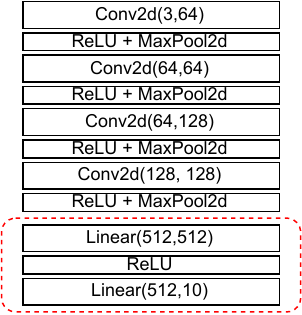}
  \caption{CNN Structure}
  \label{fig:shannon-b}
\end{subfigure}
\begin{subfigure}{0.45\linewidth}
  \includegraphics[width=\linewidth,trim={0.1em 1.8em 0.1em 0.6em},clip]
  {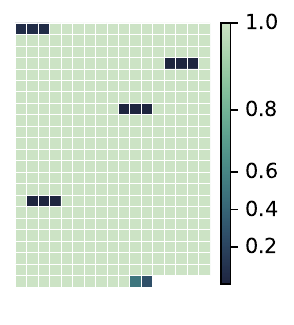}
  \caption{Weight-Vector Entropy}
  \label{fig:shannon-a}
\end{subfigure}
\caption{Heatmap~\textit{(right)} of normalized weight-vector entropies of a CNN~\textit{(left)}
containing identity-mapping kernels and RtF~(red box) primitives. Each square represents a weight 
vector. A linear layer has a single weight vector while a convolution layer has a weight vector 
for each output channel.}
\label{fig:shannon}
\end{figure}
The presence of privacy-leaking primitives during federated training is highly indicative of a 
server-mounted DRA. As a result, the feasibility of a DRA is predicated on the detectability of these 
primitives by an active client.
While \ballnumber{4} can be trivially detected as a modification to the pre-agreed architecture
by manual inspection or a structural checksum mechanism. We show that primitives~\ballnumber{1}
-- \ballnumber{3} are also easy to detect using a simple entropy-based anomaly detector. 
Given the highly specific, handcrafted nature~(identical weight rows, all zeros) of these primitives,
we expect their weight distributions to be less random when compared to layers whose weights 
have been trained organically.
To test this hypothesis, we consider a trained CNN classifier~(\autoref{fig:shannon-b}). 
We place three identity-mapping kernels~(\ballnumber{1})~(one for each input channel) in each 
convolution layer of the feature-extractor. We also replace the existing classification head 
with the RtF primitive~(\ballnumber{2}). 
We then compute the \textit{normalized entropy} of each layer's weight vector~($w$) as follows. 
First we discretize $w$ into bins of width $10^{-6}$. Treating $w$ as a draw from 
a categorical distribution, we compute the probability mass function, 
$p_i = (\text{mass of bin } \#i)/size(w)$. The normalized entropy can then be computed 
as $-\sum_i p_i\log(p_i)/\log(size(w))$. 
This value measures the dispersion in the weight distribution $w$, with $0$ entropy 
implying all elements being almost identical and $1$ implying all elements being distinct.
We present the normalized entropy of each model layer as a heatmap in \autoref{fig:shannon-a}. 
Observe, while organically trained layers show an expected high entropy~($> 0.99$), the entropies
of weight-vectors corresponding to privacy-leaking primitives are significantly lower~($< 0.5$),
and can thus be detected reliably.
We provide more results and comparisons for this mechanism in~\autoref{sec:anomaly-detection}.

%% file: sections/4-threat_model.tex
\section{Threat Model}
\label{sec:threat_model}
We consider an attacker who controls the server in the FTL setting described in~\autoref{sec:federated-learning}.
The attacker aims to reconstruct clients' private training data during federated fine-tuning. 
We assume that the server can send arbitrary model weights to clients in any round. 
However, in contrast to prior works, the server can not employ any of the privacy-leaking primitives
described in~\autoref{sec:privacy-leaking-primitives}.
\rev{Additionally, the parameters sent by the server in any round should be undetectable to the 
anomaly detection mechanism described in~\autoref{sec:anomaly-detector}.}
At the end of every round, the server receives gradient updates for \textit{only} the 
classification head and uses them to attempt data reconstruction.
The additional constraints of our threat model thus allow for a stealthier and more realistic DRA
in the FTL setting.

%% file: sections/5-methodology.tex
\section{Methodology}
\label{sec:methodology}
The feature-extractor $h_F$, receives an input sample $x$ and outputs a low-dimensional 
intermediate representation~(IR) $y$, of that sample. This IR is then used by the classification 
head $h_C$, for final classification. 
In an alternate approach to data reconstruction, Pan~\etal~\cite{pan2022exploring} present a DRA
for FL that analytically recovers IRs of the samples in a client's batch from the gradient update
and inverts them using an optimization-based approach. We refer to this process as IR-matching.
\rev{However, like gradient-matching, IR-matching also results in poor reconstructions, despite 
the use of image priors. We hypothesize that these priors are insufficient for accurate 
DRAs~(\autoref{sec:nat-image-prior}) and add \textit{robust priors} to the 
objective~(\autoref{sec:adv-prior}). This prior is imposed on $y$ using 
adversarially robust parameters for $h_F$.}
\input{sections/5-ir-matching}
\input{sections/5-ir-extraction.tex}
\input{sections/5-attack-overview.tex}

%% file: sections/5-ir-matching.tex
\subsection{DRA using Intermediate Representations}
\label{sec:ir-matching}
\begin{figure}[t!]
\centering
\begin{subfigure}{0.13\linewidth}
  \includegraphics[width=\linewidth]{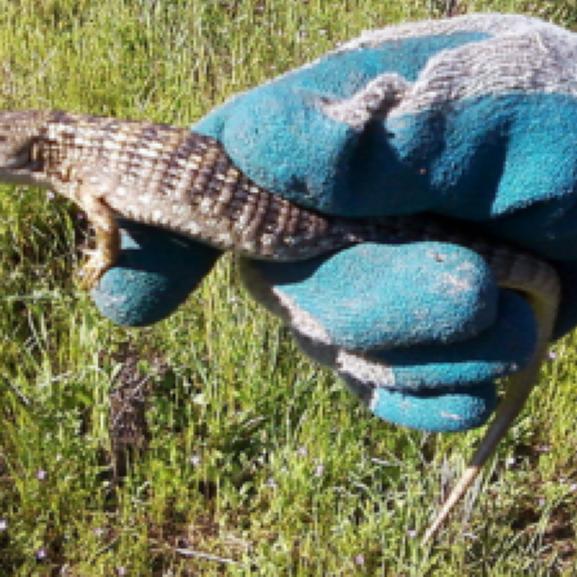}
  \caption*{$x_g$}
\end{subfigure}
\begin{tikzpicture}
  \node (y) at (0,1) {$y_g$};
  \node at (0,0) {};
  \node at (0,0) {};
  \draw[->, line width=1.1pt] (-1.0,1) -- (-0.3,1) node[midway, above] {$h_F$};
  \draw[->, line width=1.1pt] (1.0,1) -- (0.3,1) node[midway, above] {$h_F$};
\end{tikzpicture}
\begin{subfigure}{0.13\linewidth}
  \includegraphics[width=\linewidth]{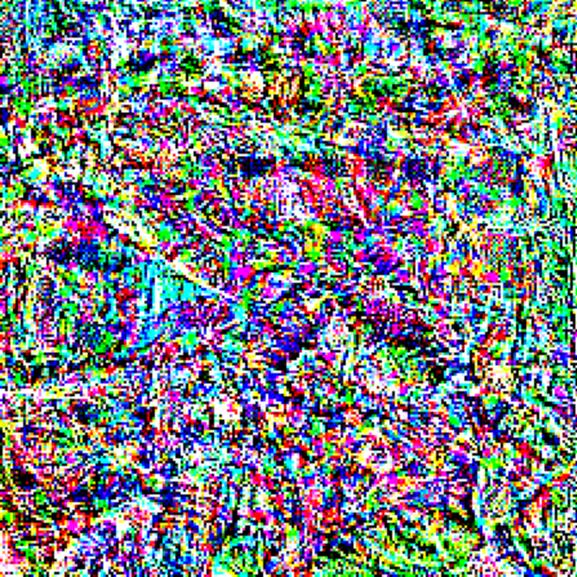}
  \caption{}
  \label{fig:artifacts-a}
\end{subfigure}
\begin{subfigure}{0.13\linewidth}
  \includegraphics[width=\linewidth]{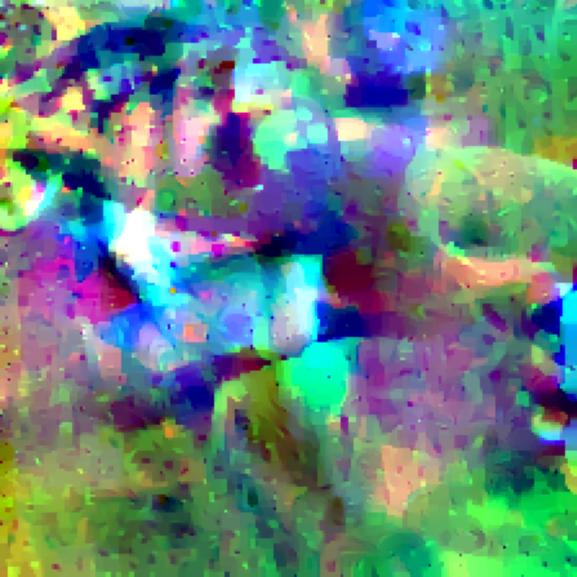}
  \caption{}
  \label{fig:artifacts-b}
\end{subfigure}
\begin{subfigure}{0.13\linewidth}
  \includegraphics[width=\linewidth]{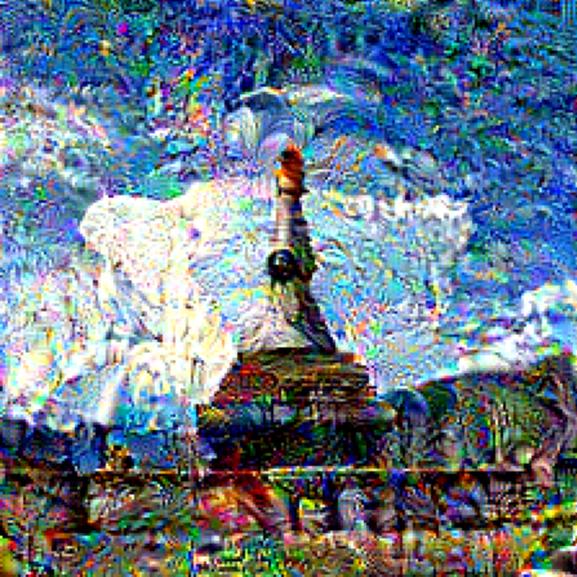}
  \caption{}
  \label{fig:artifacts-c}
\end{subfigure}
\caption{Three incorrect images reconstructed through IR-matching~($y_g$) using 
\autoref{eqn:ir-matching} in three configurations for a sample $x_g$, drawn from ImageNet.
(a) $L_2$ distance only (b) $L_2~+$ TV norm as regularizer (c) $L_2~+$ TV norm and $x$ is initialized
to an image different from the ground truth. $y_g$ is the IR generated using a trained ResNet50 
feature-extractor.}
\label{fig:artifacts}
\end{figure}
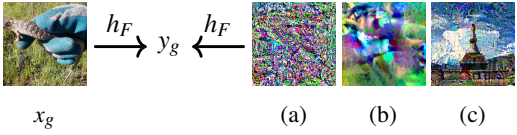
\begin{figure*}[!ht]
\centering
\begin{subfigure}{0.285\linewidth}
  \includegraphics[width=\linewidth]{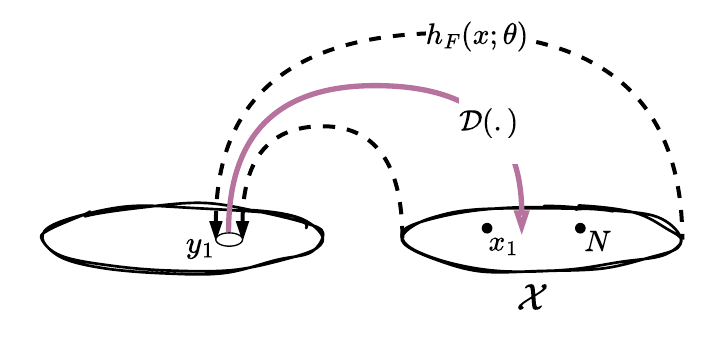}
  \caption{IR matching optimization for distance function $\mathcal D$ without adding any priors.}
  \label{fig:prior-sketch-a}
\end{subfigure}
\hspace{0.5em}
\rulesep
\hspace{0.5em}
\begin{subfigure}{0.285\linewidth}
  \includegraphics[width=\linewidth]{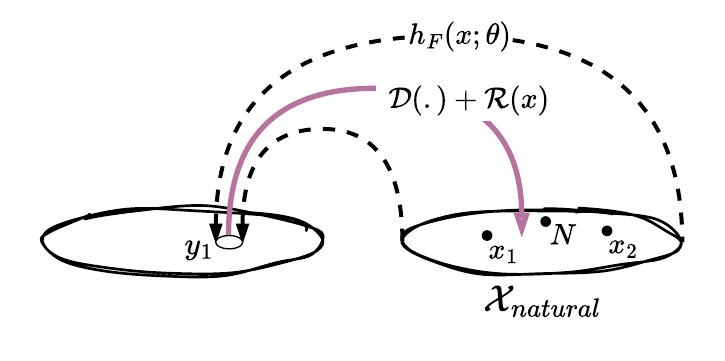}
  \caption{IR matching with natural image priors $\mathcal R$ imposed on $x$ during optimization.}
  \label{fig:prior-sketch-b}
\end{subfigure}
\hspace{0.5em}
\rulesep
\hspace{0.5em}
\begin{subfigure}{0.285\linewidth}
  \includegraphics[width=\linewidth]{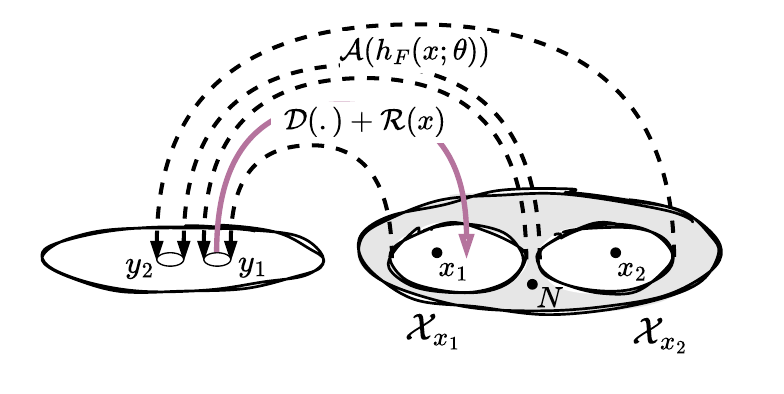}
  \caption{(Ours) IR matching with strong prior $\mathcal A$ imposed on the IR generated by $h_F$.}
  \label{fig:prior-sketch-c}
\end{subfigure}
\caption{Arrow diagrams of the mapping $x \mapsto h_F(x;\theta_F)$. \textcolor{purple}{Purple}
arrows show the inverse mapping~($h^{-1}_F$) learnt during IR matching with the action of different
priors.  (a) $\mathcal X$ is the set of preimages of $y_1$. $N$ refers to noisy samples or 
\textit{artifacts} in this region. (b) $\mathcal X_{natural}$ is the set of all \textit{natural}
images that map to $y_1$. Here, $N$ depicts artifacts that have natural image-like properties
(smooth textures, spatial locality). (c) shows the action of imposing a prior on the IRs generated 
by $h_F$ using $\mathcal A$. Here, $\mathcal X_{x_1}, \mathcal X_{x_2}$ is the set of natural images
resembling $x_1$ and $x_2$ respectively. 
Note, $x_1, x_2$ do not belong to the preimage sets of $y_2, y_1$ respectively.}
\label{fig:prior-sketch}
\end{figure*}

The goal of IR-matching is to learn the inverse mapping $h_F^{-1}$ given $\theta_F$. 
Consider a single sample-IR pair $(x_g, y_g)$ such that $y_g = h_F(x_g;\theta_F)$. 
Then, similar to~\autoref{eqn:gradient-matching}, IR-matching is 
{
\begin{gather}
\label{eqn:ir-matching}
x^* = \argmin_x \Big[ \mathcal D(h_F(x; \theta_F ), y_g) + \mathcal R(x)\Big]
\end{gather}
}%
The value of $x$ is iteratively refined, starting from a random guess, to minimize the $\mathcal D$-distance
between the generated IR $h_F(x;\theta_F)$ and $y_g$. 
Unfortunately, like gradient-matching, \autoref{eqn:ir-matching} tends to converge to local optimas
for $x$ when using typical distance functions like $L_1,L_2$ or cosine distance. 
This results in the generation of \textit{``artifacts''}, i.e., images that minimize the IR-matching
loss but bear no perceptual resemblance to the ground truth.
\autoref{fig:artifacts} presents three very distinct preimages for the IR of $x_g$~(\ie $y_g$) 
each having a small $L_2$ IR-matching distance~($<10^{-4}$).

This tendency highlights the fundamentally \emph{many-to-one} nature of the $h_F$ mapping.
To add tighter constraints on the search space, DRAs need to rely heavily on prior knowledge of 
the sample's domain~\cite{balunovic2022bayesian}.
Adding a regularization term $\mathcal R(x)$, to the objective allows for domain specific priors
to be incorporated into the optimization process.
Existing works have used TV norm~\cite{geiping2020inverting}, $L_2$ norm~\cite{yin2021see}, 
DeepInversion~\cite{yin2021see}, GC~\cite{yin2021see} and DIP~\cite{pan2022exploring}, to guide
the optimization towards ``natural'' images.
We refer to these as \textit{natural image priors}.

\subsubsection{Insufficiency of Natural Image Priors}
\label{sec:nat-image-prior}
Consider two randomly sampled distinct natural images $x_1, x_2$. 
Engstrom~\etal~\cite{engstrom2019adversarial} show that $x_2$ can be minimally perturbed, such 
that the IRs of $x_1, x_2$ are nearly identical. The perturbation $\delta$, which minimizes the 
$L_2$ IR-matching distance can be obtained by solving the following objective -- 
{\small
\begin{align}
\label{eqn:pert-examples}
\argmin_{\delta \in \Delta} \Big[ 
  \big\lVert h_F(x_1; \theta_F) - h_F(x_2+\delta; \theta_F)\big\rVert^2_2 + \mathcal R(x_2+\delta)
\Big] 
\end{align}
}%
where $\Delta$ is the set of allowed~(small) perturbations. 
One such preimage is visualized in~\autoref{fig:artifacts-c} for a trained ResNet50 
feature-extractor. 
By solving~\autoref{eqn:pert-examples}, multiple preimages in the domain of natural images can
be found for any given IR. Each of these preimages are viable solutions to the IR-matching problem
and can potentially be reached by the optimizer. 
This idea is sketched out in \autoref{fig:prior-sketch}. During unregularized IR-matching
(\autoref{fig:prior-sketch-a}), the optimizer can converge to any image in the preimage set of 
$y_1$~(depicted by $\mathcal X$) which also contains artifacts~($N$). 
With the addition of natural image priors~(\autoref{fig:prior-sketch-b}), the preimage set~($\mathcal X_{natural}$)
contains natural looking images~($x_2$) and artifacts which conform to the prior
(for example, having smooth textures as in~\autoref{fig:artifacts-b}), that are very different from $x_1$.
Based on this we argue that relying solely on natural image priors is insufficient for IR-matching.

Unlike prior DRAs which have imposed various priors on the reconstruction candidate, 
our key idea is to impose an additional, strong prior on the IR itself to strictly constrain the
preimage space, thereby  improving reconstruction accuracy. 

\subsubsection{Robustness as a Strong Prior for IR-Matching}
\label{sec:adv-prior}
The motivation for imposing a prior distribution on IRs is to ensure that well separated inputs
in the image space have well-separated representations in the IR space. 
This tightly constrains the search space for IR-matching to a set of images that 
closely resemble the original sample~(\autoref{fig:prior-sketch-c}). 
We impose such a prior through adversarial training~($\mathcal A$) of $h_F$~\cite{engstrom2019adversarial}.

\noindent\textbf{Adversarial Training~(AT).} 
AT is used to train a classifier whose predictions are invariant under minor perturbations to 
the input. Specifically, AT learns robust parameters $\hat\theta$ such that the predicted category
of $h(x; \hat\theta) = h(x + \delta; \hat\theta)~ \forall \delta \in \Delta$, where $\Delta$ is
the set of allowed~(small) perturbations. 
AT is done by minimizing the expected maximum loss $L$ over $\Delta$~(\cite{madry2018towards}) as --
{
\begin{align}
\label{eqn:AT}
\hat\theta= \argmin_\theta \mathbb E_{(x,c) \sim D}\Big[\max_{\delta \in \Delta} L(h(x+\delta; \theta), c) \Big]
\end{align}
}%
AT trains a feature-extractor to generate IRs invariant to small perturbations in the 
input~\cite{engstrom2019adversarial}. The distribution of these \textit{robust-IRs} shows a strong
correlation with high-level, semantic features present in the input. This results in distinct 
natural images with distinct features having well separated IRs. 
AT is thus said to impose a ``robust prior'' on the IR distribution. 

\noindent\textbf{Data Reconstruction using Robust IRs.}
\label{sec:robust-ir-matching}
Robust IRs can be inverted back to their corresponding input samples by solving~\autoref{eqn:ir-matching}. 
To further improve the reconstruction accuracy, we also incorporate the DIP~\cite{ulyanov2018deep}~(similar 
to~\cite{pan2022exploring}) during optimization. To do so, we employ a generator network $h_G(s;\theta_G)$ 
which maps a latent vector $s$ to image $x$. 
Then, given a robust-IR $\hat y = h_F(x_g;\hat\theta_F)$ of some sample $x_g$, we can solve --
{
\begin{gather}
\label{eqn:robust-ir-matching}
x^* = h_G(s^*; \theta_G^*) ~~~~~ \text{such that}\\
s^*,\theta_G^*  = \argmin_{s,\theta_G} \Big[
  \mathcal D(h_F(h_G(s; \theta_G); \hat \theta_F), \hat y) + \mathcal R(h_G(s;\theta_G))
  \Big] \nonumber
\end{gather}
}%
We use a combination of KL divergence and MSE loss for the distance metric $\mathcal D$. For $\mathcal R$, we simply use the TV norm.
The detailed IR matching algorithm is presented in~\autoref{alg:recon}. 
\input{sections/inversion-algo.tex}
The hyperparameter $\alpha$ is used to weigh the contribution of both losses. 
The KL divergence loss captures the robust-IR's inter-feature dependencies and guides the 
optimization towards an image whose IR has a similar joint distribution. 
The MSE loss on the other hand encourages independent similarity of each feature of the generated IR
with the robust-IR.
To avoid getting stuck in local minimas and generating artifacts, 
we also periodically add gaussian noise to the seed during optimization.
More details about the generator architecture and hyperparameters are presented subsequently 
in~\autoref{sec:implementation}.

%% file: sections/inversion-algo.tex
\begin{algorithm}[t!]
\caption{IR-Matching Algorithm}
\label{alg:recon}
\small
\begin{algorithmic}[1]
  \Require Robust IR: $\hat y$, FE: $h_F(.;\hat\theta_F)$, Generator: $h_G(.;\theta_G)$, 
  Iterations: $K$, Step size: $\eta_1, \eta_2$, Regularization parameters: $\alpha, \beta, K_1$
  \State $s \gets \Phi(\mathbf 0, \mathbf 1)$ \Comment{Sample $s$ from a standard normal distribution}
  \For{$k=1$ to $K$}
  \If{$k \mod K_1 = 0$}
      \State $s \gets s + \Phi(\mathbf 0,\mathbf 1)$ \Comment{Perturb $s$ every $K_1$ iterations}
  \EndIf
  \State $x \gets h_G(s;\theta_G)$
  \State $y \gets h_F(x; \hat\theta_F)$
  \State $L \gets \alpha\cdot\text{KL}(y, \hat y) + (1-\alpha)\text{MSE}(y,\hat y) +  \beta\cdot\text{TV}(x)$
  \State $s \gets s - \eta_1 \nabla_s L$
  \State $\theta_G \gets \theta_G  - \eta_2 \nabla_{\theta_G}  L$
  \EndFor
  \Ensure Reconstructed sample $x^* \gets h_G(s; \theta_G)$
\end{algorithmic}
\end{algorithm}

%% file: sections/5-ir-extraction.tex
\subsection{Extracting IRs from the Gradient Update}
\label{sec:ir-extraction}
During federated finetuning, the server simply receives gradients of the classification head
$h_C$ and does not have direct access to the IRs of a client's batch. 
These IRs need to be extracted from the gradient update itself.
It is known that under certain conditions, the batched input received by a linear layer can be 
recovered from the layer's weight and bias gradients~\cite{pan2022exploring,zhu2020rgap}.
This mechanism is known as {\em linear layer leakage}. 
Thus, a malicious server can extract IRs of the client's batch from the gradients of the linear 
layer directly following the feature-extractor~(or equivalently, at the beginning of $h_C$),
when the conditions of linear layer leakage are satisfied.

\subsubsection{Linear Layer Leakage}
\noindent\textbf{Notation.}
We use uppercase letters in boldface~($\mathbf X$) for matrices representing a batch of real 
vectors~(represented by lowercase letters) arranged in rows. 
The $i^{th}$ row or column of $\mathbf X$ is represented in shorthand as $\mathbf X(i,:)$ and 
$\mathbf X(:,i)$ respectively.
\input{figures/spab_blk.tex}

\vspace*{-1em}
Consider a linear layer parameterized by the weight and bias vectors $w \in \mathbb R^{M \times N}$ 
and $b \in \mathbb R^{N}$ respectively. This layer receives a $B$ sized batch of 
$M$ dimensional IRs $\mathbf Y \in \mathbb R^{B \times M}$ and produces a batched output 
$\mathbf Z = \mathbf{Y}w + b$. 
If a scalar loss is computed on $\mathbf Z$, the weight and bias gradients can be written in 
terms of the output gradient $\nabla \mathbf Z$ as follows --
{
$$
\nabla w(i,j) = \sum_{k=1}^{B}\mathbf Y(k,i) \nabla \mathbf Z(k,j), \hspace*{1em} 
\nabla  b(j) = \sum_{k=1}^{B} \nabla \mathbf Z(k,j)
$$
}%
In general, this linear system does not have a unique closed-form solution. 
However, observe that if $\nabla \mathbf Z$ has a \textit{single nonzero element} in say, 
column $q$ and row $p$, then the isolated IR $\mathbf Y(p,:)$ can be precisely recovered as -- 
{
\begin{align}
\label{eqn:ir-extraction}
\mathbf Y(p,:) = \frac{\nabla w(:,q)}{\nabla b(q)}  \hspace{1em} \text{where}
\hspace{1em} \nabla \mathbf Z(p,q)~,~\nabla b(q)\neq 0 
\end{align}
}%
\highlightbox{\textbf{Observation 1. }If $\nabla \mathbf Z$ is $N$ column-sparse, \ie each 
column has exactly one non-zero element, then atmost $N$ IRs~(rows of $\mathbf Y$) can be 
extracted from $\nabla \mathbf Z$.}

\noindent\textbf{IR Leakage Rate.} 
It is defined as the fraction of the client batch whose IRs can be extracted precisely from the
gradient update.
It is equal to the number of sparse columns in $\nabla \mathbf Z$ that have non-zero elements 
in \underline{distinct} rows.

\noindent\textbf{ReLU-Induced Column Sparsity.}
If a ReLU activation is applied to the output of the linear layer as,
$\mathbf Z' = \text{ReLU}(\mathbf Z)$ then, the column sparsity of $\nabla \mathbf Z$ is 
affected as follows.
{
\begin{align}
 \nabla \mathbf Z(i,j)  =
 \begin{cases}
     \nabla \mathbf Z'(i,j)  &\mathbf Z(i,j) > 0\\
     0 &\text{otherwise}
 \end{cases} 
\end{align}
}%
\begin{figure*}[t!]
  \centering
  \begin{subfigure}{\linewidth}
    \centering
    \includegraphics[width=0.9\linewidth, trim={0em 1.5em 0em 0em}, clip]{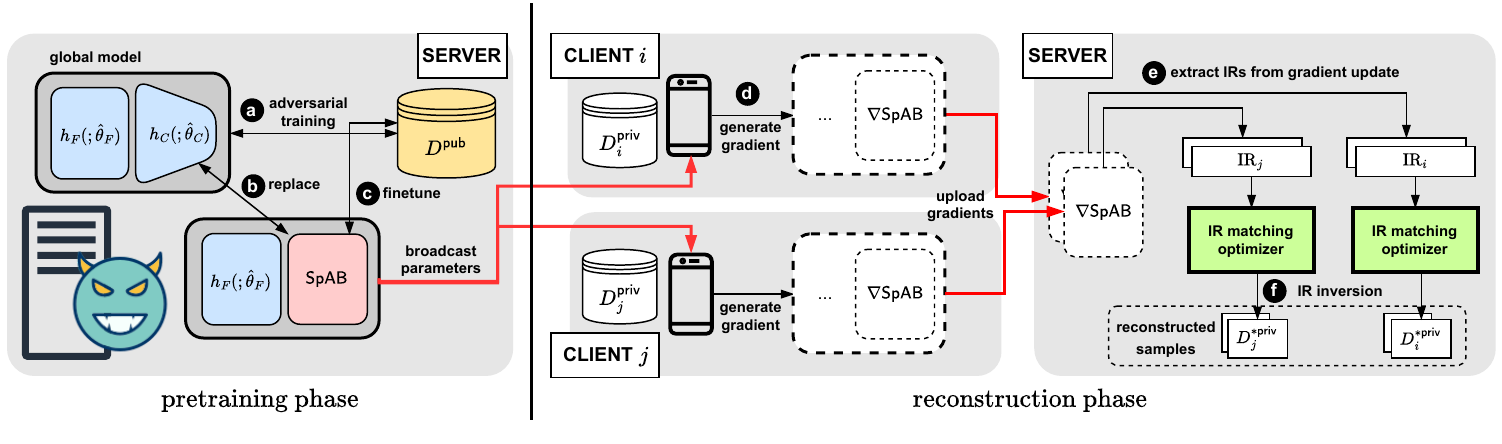}
  \end{subfigure}
  \caption{Overview of \abc, a DRA specifically targeted to Federated Transfer Learning (FTL).}
  \label{fig:overview-fig}
\end{figure*}
%
\highlightbox{\textbf{Observation 2. }For a ReLU activated linear layer, the column sparsity of
the output gradient $\nabla \mathbf Z$ can be controlled by the column sparsity of activation 
$\mathbf Z'$, if the gradient flowing back~($\nabla \mathbf Z'$) is non zero.
}%
We find that the linear layers present in a trained model's classification
head typically exhibit a low IR leakage rate~(\autoref{sec:evaluation}).
This bottlenecks the efficiency of an IR-based DRA. 
It arises because of a lack of column sparsity in the output gradient. 
\rev{RtF~\cite{fowl2022robbing} propose an architecture and parameter distribution for $h_C$ which 
significantly improve the leakage rate. However, as discussed in~\autoref{sec:privacy-leaking-primitives},
these are detectable by an active client.
Instead, we propose \textit{learning} $h_C$ parameters to improve leakage rate, while respecting
the constraints of our threat model.
}

\subsubsection{Improving IR Leakage through \spab-Training}
\label{sec:spab-loss}
To leak IRs from a model, we replace its classification head with a 2-layer, ReLU activated MLP as 
shown in~\autoref{fig:spab-blk}.
We refer to this module as the \textit{sparse activation block}~(\spab). The first linear layer is
parameterized by weight, bias vectors $w \in \mathbb R^{M\times N}, b \in \mathbb R^N$ respectively.
This is represented by the shorthand notation~$\spab(N)$.
The aim is to train the \spab module such that the first linear layer has a large IR leakage rate.
The key idea is to achieve this by encouraging column-sparsity in the activation
$\mathbf Z'$ during standard classifier training~(\textit{Observation 2}).
In ML literature, $L_1$-norm minimization is a well-known technique for learning sparse weights
(for instance, during Lasso regularization). Similarly, we introduce a regularization term, 
$ \frac{1}{N}\sum_{j=1}^N \lVert \mathbf Z'(:, j) \rVert_1$ which minimizes the $L_1$-norm of its 
\textit{columns} to induce column-sparsity.
Here, to avoid the trivial $\mathbf Z'=0$ solution we consider an additional loss term 
${\frac{1}{BN}\sum_{i=1}^{B}\sum_{j=1}^{N} \max(0, -\mathbf Z(i, j))}$
to penalize negative values in $\mathbf Z$ and encourage positive activations.
Instead of directly adding this term, we first relax the max operator with the softplus approximation
to get a smooth loss.
We refer to the final regularization term as the sparsity loss $L_\text{sp}$
{
\begin{gather}
  L_\text{sp} :=\frac{\beta_1 }{N}\sum_{j=1}^N \lVert \mathbf Z'(:, j) \rVert_1 
  + {\frac{\beta_2}{BN}\sum_{i=1}^{B}\sum_{j=1}^{N} \log(1+e^{-\mathbf Z(i, j)})} \nonumber
\end{gather}
}%
Adding $L_\text{sp}$ to the classification loss $L_\text{cls}$, we get the final \spab-training
objective for learning parameters which result in large scale IR leakage from the \spab module.
{
\begin{gather}
  \theta^\textsf{spab}_C = \argmin_{\theta_C} \big[ \alpha L_\text{cls} + (1-\alpha)L_\text{sp} \big]
  \label{eqn:final-loss}
\end{gather}
}%
$\alpha, \beta_1, \beta_2$ are hyperparameters used to stabilize training. 
Note that $L_\text{cls}$~(computed as the cross entropy loss) is minimized over all parameters of
$h_C$ which includes $w, b$.
We refer to this process as \spab-training and present a detailed algorithm in~\autoref{alg:spab-training}.
\input{sections/spab-algo.tex}
Given a model $h$, \spab-training first replaces the existing classification head with a 
$\spab(\kappa)$ block. 
The value of $\kappa$ is a hyperparameter and can be used to manipulate the leakage rate with 
larger values typically allowing higher leakage rates.
The bias vector $b$ is initialized to $0$. 
The hyperparameter $\alpha$ balances the strength of $L_\text{sp}$ regularization and is varied
according to a cosine schedule. 
The parameters $w, b$ are also perturbed with noise in each epoch. 
Experimentally, we found this step to additionally improve the IR leakage rate.

\noindent\textbf{Post training IR recovery.}
Essentially, \spab-training trains a linear layer such that it generates column-sparse activations for 
a batch of inputs. As a result, IRs input to this layer can easily be extracted from its gradients
by using \autoref{eqn:ir-extraction}.

%% file: figures/spab_blk.tex
\begin{figure}[t!] 
\centering
\resizebox{0.9\linewidth}{!}{%
\begin{tikzpicture}[
  block/.style={
    rectangle, 
    draw, 
    thick,
    text centered, 
    align=center,
    fill=white
  },
  connector/.style={-Latex,very thick}
 ]
  \node[block, minimum height=1.5cm, minimum width=1.0cm] (rect1) {$h_F(;\hat\theta_F)$};
  \node[block, right=1.4cm of rect1, minimum height=1.5cm, minimum width=1.0cm] (rect2) 
  {$w_{M\times N}$\\ \\$b_{N}$};
  \node[block, right=1.0cm of rect2, minimum height=1.5cm, minimum width=0.6cm, label={[rotate=90]center:ReLU}] (rect3) {};
  \node[block, right=1.0cm of rect3, minimum height=1.5cm, minimum width=1cm] (rect4) 
  {$w'_{N\times C}$\\ \\$b'_{C}$};
  
  \coordinate[left=1.0cm of rect1.west] (input);
  \draw[connector] (input) -- node[above, midway] {$\mathbf X_{B\times ...}$} (rect1.west);
  \draw[connector] (rect1.east) -- node[above, midway] {$\mathbf Y_{B\times M}$} (rect2.west);
  \draw[connector] (rect2.east) -- node[above, midway] {$\mathbf Z_{B\times N}$} (rect3.west);
  \draw[connector] (rect3.east) -- node[above, midway] {$\mathbf Z'_{B\times N}$} (rect4.west);

  \begin{pgfonlayer}{background}
    \node[
      draw, 
      dashed,
      thick, 
      rounded corners=10pt, 
      fill=gray!30, 
      fit={(rect2) (rect3) (rect4)}, 
      inner sep=0.2cm, 
      label={[draw, fill=white, thick, yshift=-1mm]north:{\small$\spab(N)$}} 
    ] (container) {};
  \end{pgfonlayer}

  \node[below=0.25cm of rect1] (caption1) {\textbf{\text{feature extractor}}};
  \node[below=0.05cm of container] (caption_container) {\textbf{\text{classification head}}};

\end{tikzpicture}%
}
\caption{Placement and structure of the $\spab(N$) construction. IRs~($\mathbf Y$) of 
$\mathbf X$ are recovered from the gradients of $w, b$.}
\label{fig:spab-blk}
\end{figure}

%% file: sections/spab-algo.tex
\begin{algorithm}[t!]
\caption{\spab Training Algorithm}
\label{alg:spab-training}
\small
\begin{algorithmic}[1]
  \Require Model: $h_C(h_F(.;\hat\theta_F);\theta_C)$, Epoch: $K$,  Dataset: $D^\textsf{pub}$, Step size: $\eta$, \spab size: $\kappa$,
  Hyperparameters: $\beta_1, \beta_2,\sigma$
  \State $h_C \gets \spab(\kappa)$ \Comment{replace $h_C$ with $\spab(\kappa)$}
  \State $b \gets \mathbf 0$ \Comment{initialize bias vector to 0}
  \For{$k=1$ to $K$}
  \State $\alpha \gets 1-\cos^2(k\pi/2K)$
  \State $\{w,b\} \gets \{w,b\} + \Phi(0,\sigma^2)$ \Comment{perturb $w, b$}
  \For{$(\mathbf X, \mathbf C) \sim D^\textsf{pub}$} \Comment{training with batch size $B$}
    \State \textbf{\footnotesize acquire activations $\mathbf Z, \mathbf Z'$ during forward pass}
    \State {\small$L_\text{cls} = \text{CELoss}(h_C(h_F(\mathbf X; \hat\theta_F);\theta_C), \mathbf C)$}
    \State \textbf{\footnotesize compute sparsity loss using activations}
    \State {\small $L_\text{sp} = \frac{\beta_1}{N}\sum_{j}^{} \lVert \mathbf Z'(:, j)\rVert_1 + {\frac{\beta_2}{BN}\sum_{i}^{}\sum_{j}^{}\log(1+e^{-\mathbf Z(i, j)})}$}
    \State \textbf{\footnotesize compute total loss and descend}
    \State $L = \alpha L_\text{cls} + (1-\alpha)L_\text{sp}$
    \State $\theta_C \gets \theta_C - \eta \nabla_{\theta_C} L$

  \EndFor
  \EndFor
  \Ensure Final Model $h_C(h_F(; \hat\theta_F);\theta^\textsf{spab}_C)$
\end{algorithmic}
\end{algorithm}

%% file: sections/5-attack-overview.tex
\subsection{Attack Mechanism}
\label{sec:overview}
We now present our DRA -- \abc, which leverages robust priors and \spab-training to mount data
reconstruction during federated finetuning. 
\abc operates in two phases --- 
(i) the {\textit{pretraining phase}} which coincides with the centralized pretraining of the 
global model for transfer learning, and 
(ii) the {\textit{reconstruction phase}} which occurs during federated finetuning. In this phase
the server receives a gradient update, extracts robust IRs and performs IR-matching to reconstruct
private inputs. The two stages are illustrated in~\autoref{fig:overview-fig} and a detailed 
description of each stage is presented below --

\noindent\textbf{Pretraining Phase.}
In this phase, the server uses $D^\textsf{pub}$ to adversarially train the global model~(\ballnumber{a}).
The aim is to learn a robust feature-extractor $h_F(\cdot;\hat\theta_F)$ to impose a robust prior
on the generated IRs.
Once trained, the server replaces the classification head with a \spab module~(\ballnumber{b}).
The feature extractor is frozen and the server performs \spab-training on $D^\textsf{pub}$~(\ballnumber{c}).
The feature extractor parameters $\hat\theta_F$ are then broadcast to all clients.
Here, we note that replacing the classification head with a \spab module does not violate our
threat model. This is further discussed in detail in~\autoref{sec:discussion}.  

\noindent\textbf{Reconstruction Phase.} 
During federated finetuning, in any chosen round, the server broadcasts parameters $\theta^\textsf{spab}_C$
computed during the pretraining phase to all clients and receives the classification head~(or 
equivalently \spab module) gradients. 
Robust IRs corresponding to training samples can then be extracted from each gradient 
using~\autoref{eqn:ir-extraction}~(\ballnumber{e}). 
Each IR is then inverted to its corresponding sample by IR-matching~(\autoref{sec:robust-ir-matching})
(\ballnumber{f}). Note that the server can perform  IR-matching parallely for each extracted IR.

%% file: sections/6-implementation.tex
\section{Evaluation Setup and Implementation}
\label{sec:evaluation-details}
We consider a single client--server interaction for our evaluations. 
The client samples a batch at random from its private dataset and computes the gradients of the
classification head using FedSGD and sends it to the server. 
The server extracts IRs from the gradient and performs IR-matching to reconstruct the private sample
corresponding to each extracted IR.

\subsection{DRA Performance Metrics}
We evaluate \abc's performance on the basis of data reconstruction accuracy and efficiency.
Reconstruction accuracy measures visual similarity between the reconstructed and original image
sample. For this, we consider an ensemble of different commonly used metrics like PSNR, SSIM 
and LPIPS scores~(more information on these metrics is provided in~\autoref{app:reconstruction-metrics}).
However, we note that no single metric captures true human-level perceptual quality. 
Therefore, we also present several visual comparisons as qualitative evidence.

Reconstruction efficiency is the fraction of a client's batch reconstructed with sufficient accuracy.
For this, we use the IR leakage rate~(defined in~\autoref{sec:ir-extraction}).
Formally, it is defined as the ratio $(\textit{\# of distinct IRs extracted from \spab-gradient})/B$.

\subsection{Datasets and Model Architectures}
\begin{table}[t!]
\caption{Summary of datasets and models used for evaluations. Feature extractor size is in 
terms of number of trainable parameters. NAcc and RAcc refer to the natural and robust accuracy
achieved through AT on the corresponding datasets.}
\centering
\resizebox{0.9\columnwidth}{!}{%
\renewcommand{\arraystretch}{1.0}
\begin{tabular}{ccccll}
\hline
\textbf{Dataset} & \textbf{Model} & \textbf{\begin{tabular}[c]{@{}c@{}}$h_F$ Size\\ (\# Params)\end{tabular}} & \textbf{\begin{tabular}[c]{@{}c@{}}IR\\ Size\end{tabular}} & \textbf{NAcc} & \textbf{RAcc} \\ \hline
\multirow{3}{*}{\begin{tabular}[c]{@{}c@{}}CIFAR10\\ $3\times 32\times 32$\end{tabular}} & CNN & 260K & 512 & 0.72 & 0.51 \\ \cline{2-6} 
 & ResNet18 & 11.1 M & 512 & 0.84 & 0.60 \\ \cline{2-6} 
 & ShuffleNet-V2 & 341 K & 1024 & 0.78 & 0.55 \\ \hline
\multirow{3}{*}{\begin{tabular}[c]{@{}c@{}}ImageNet\\ $3\times 224 \times 224$\end{tabular}} & VGG-11 BN & 9.2 M & 25088 & 0.52  & 0.29 \\ \cline{2-6} 
 & ResNet50 & 23 M & 2048 & 0.61 & 0.37 \\ \cline{2-6} 
 & ViT B-32 & 87 M & 768 & 0.49 & 0.24 \\ \hline
\end{tabular}%
}
\label{tab:eval-summary}
\end{table}
A summary of the main datasets and models used is presented in~\autoref{tab:eval-summary}.
For our main evaluations, we consider two datasets, CIFAR10~\cite{krizhevsky2009learning} and 
ImageNet~\cite{russakovsky2015imagenet}.
The training split of each dataset is assumed to be $D^\textsf{pub}$ while the validation/test
split is taken as $D^\textsf{priv}$.
We consider three shallow architectures for CIFAR10 -- CNN, ResNet18, ShuffleNet-V2 and three 
complex architectures for ImageNet -- ResNet50, VGG11 with BatchNorm~(VGG 11-BN) and a 
Vision Transformer~(ViT B-32). 
For the ImageNet models, we use implementations from the torchvision library. For CIFAR10 models,
we use open-source implementations\cite{kuangliu} of ResNet18 and ShuffleNet-V2. The CNN architecture
is presented in~\autoref{fig:shannon-b}. 
We additionally use three datasets-RetinaMNIST~\cite{medmnistv1}, BloodMNIST~\cite{medmnistv1}
and CelebA~\cite{liu2015faceattributes} for out-of-distribution  
experiments~(\autoref{sec:ood-recon}).

\subsection{Implementation and Training Details}
\label{sec:implementation}
We developed the proof-of-concepts for this work using PyTorch 2. All experiments were run on 
an Nvidia A100-80GB GPU. The IR-matching is computationally intensive and requires a moderate 
amount of GPU memory depending on the model, generator architecture and input sample size. 
Our codebase can be found here\footnote{\url{https://github.com/theboxahaan/maui}}.
We now present details on the various training protocols used in \abc's pretraining phase

\noindent\textbf{Adversarial Training.}
For AT, we use automatic mixed precision~(AMP) with gradient scaling for speeding up AT on the 
GPUs~\cite{torch2025amp}.
All models are trained using SGD with momentum and a step scheduler having a decay factor and 
period of $0.1, 50$ respectively. The initial learning rate is set to $0.128$. We use the Cutout 
data-augmentation technique~\cite{rebuffi2021data} to further boost robustness.
Adversarial samples are generated using $L_\infty$-norm PGD with a perturbation budget of ${4}/{255}$.
The ImageNet models are trained with a batch size of $512$ for $90$ epochs and the model with 
the highest robustness is selected. The PGD step size is $4/255\cdot2/3$ and number of steps taken
is $3$.
For the CIFAR10 models, the batch size is $128$ and the PGD parameters are step-size of $2/255$
with $10$ steps. The natural, robust accuracies of the models are presented 
in~\autoref{tab:eval-summary}.

\begin{figure}[t!]
  \centering
  \begin{subfigure}{0.44\linewidth}
    \includegraphics[width=\linewidth,trim={0 0em 0em 0}, clip]{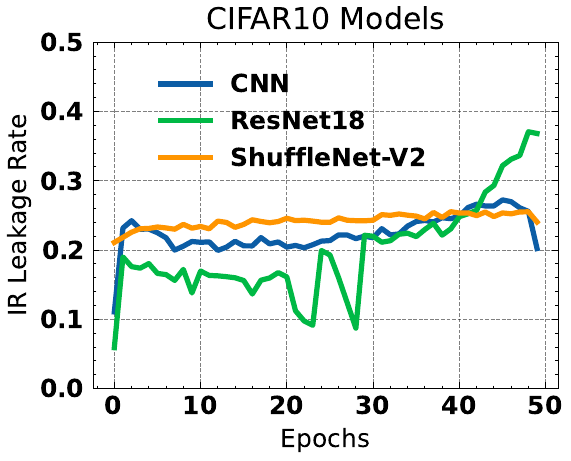}
  \end{subfigure}
  \begin{subfigure}{0.44\linewidth}
    \includegraphics[width=\linewidth]{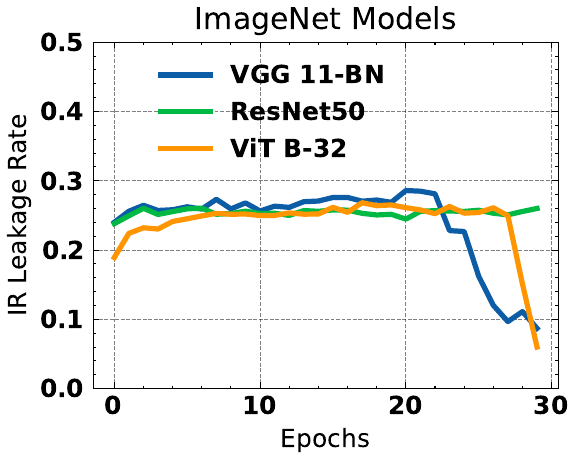}
  \end{subfigure}
  \caption{Leakage rate trends during \spab training.}
  \label{fig:finetuning-trends}
\end{figure}

\noindent\textbf{\spab-Training.}
We replace the classification heads of all models with a $\spab(512)$ block for \spab-training.
The feature-extractor is frozen and only the \spab module is trained according to~\autoref{alg:spab-training}
with a batch size of $64$ and the Adam optimizer. 
To speed up training, we only iterate over $400$ batches per epoch.
The ImageNet and CIFAR10 models are trained for 30 and 50 epochs respectively.
We tune the \spab hyperparameters and learning rate using a Tree-Parzen Estimator from the hyperopt
library~\cite{bergstra2013making}. The final values are presented in~\autoref{tab:spab-conf} 
and their corresponding training curves in~\autoref{fig:finetuning-trends}. 
The leakage rate is computed on $D^\textsf{priv}$ for $B=512$. 
\begin{table}[t!]
\centering
\caption{Final $\spab(512)$-training hyperparameters.}
\label{tab:spab-conf}
\resizebox{0.9\columnwidth}{!}{%
\begin{tabular}{cccccc}
\hline
\textbf{Model} & \textbf{\begin{tabular}[c]{@{}c@{}}IR leakage\\ Rate$(B=512)$\end{tabular}} & \textbf{$\eta$} & \textbf{$\beta_1$} & \textbf{$\beta_2$} & \textbf{$\sigma$} \\ \hline
CNN           & $0.26$ & $0.0075$ & $17868.661$ & $63.2169$ & $0.0225$ \\ \hline
ResNet18      & $0.37$ & $0.0328$ & $17140.799$ & $66.6352$ & $0.0863$ \\ \hline
ShuffleNet-V2 & $0.26$ & $0.0015$ & $19538.475$ & $50.9420$ & $0.1320$ \\ \hline
VGG 11-BN     & $0.28$ & $0.0022$ & $19593.060$ & $52.3568$ & $0.0867$ \\ \hline
ResNet50      & $0.28$ & $0.0009$ & $13481.506$ & $44.7335$ & $0.0716$ \\ \hline
ViT-B32       & $0.27$ & $0.0019$ & $12794.071$ & $53.0163$ & $0.0043$ \\ \hline
\end{tabular}%
}
\end{table}%
\noindent\textbf{IR Matching.}
For the generator $h_G$, we use a U-net style network which operates on a latent vector of the 
same dimensions as the data sample.
For improved performance we compile the generator and use medium precision bfloat16 matrix multiplications.
We use the AdamW optimizer with a step-LR scheduler and run the optimization loop for 
$30,000$ and $8,000$ iterations for ImageNet and CIFAR10 models respectively.

%% file: sections/7-evaluation.tex
\section{Evaluations and Results}
\label{sec:evaluation}
\subsection{Generating Natural Preimages}
\label{sec:colliding-samples}
\begin{figure}[t!]
\centering
\begin{subfigure}{0.9\linewidth}
  \includegraphics[width=\linewidth,trim={0.1em 2.2em 0.1em 0em}, clip]{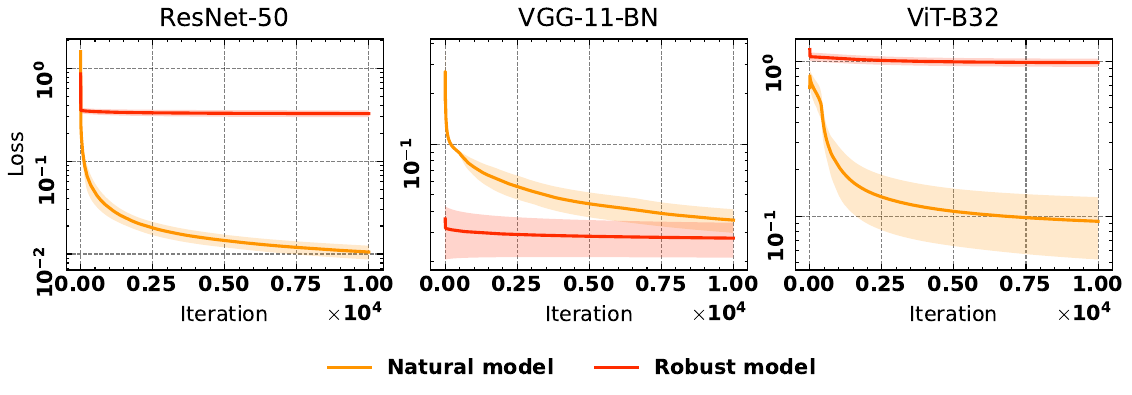}
\end{subfigure}
\begin{subfigure}{0.9\linewidth}
  \includegraphics[width=\linewidth,trim={0.1em 0 0.1em 0}, clip]{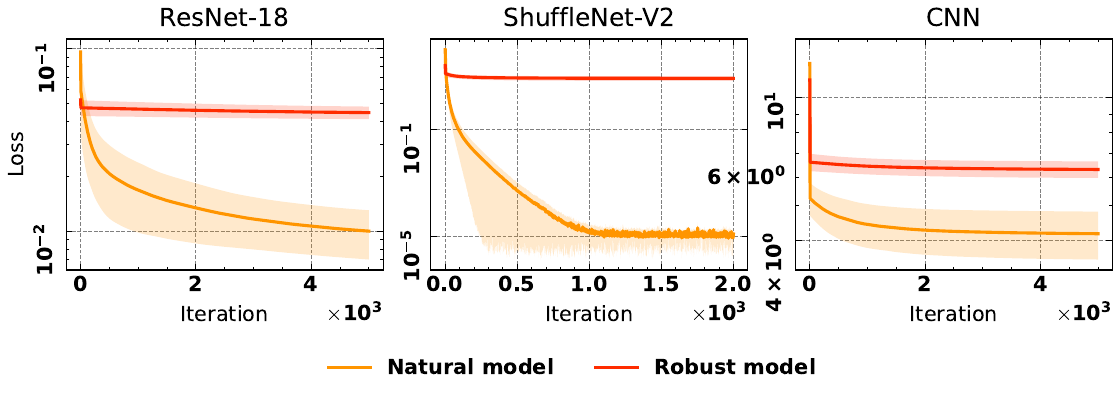}
\end{subfigure}
\caption{Distance between IRs of $x_1, x_2$ on naturally trained and robust feature-extractor
(i.e.,\abc) of different models.}
\label{fig:ir-distance}
\end{figure}
We empirically validate the claim that natural priors are insufficient for accurate 
IR-matching~(\autoref{sec:nat-image-prior}) by generating distinct multiple natural-looking 
preimages for a given IR.
We randomly draw an image $x_1$ and select three distinct images at random as seeds for 
three preimages $\{x_2\}$'s.
We use TV-norm as the natural image prior.
To each $x_2$, we add minute perturbations constrained to an $L_\infty$-ball of radius $4/255$
centered at $x_2$ and iteratively refine these perturbations such that the generated IR matches
the IR of $x_1$~(\autoref{eqn:pert-examples}).
\autoref{fig:ir-distance} shows the evolution of the IR-matching distance during optimization 
on different naturally trained feature-extractors. 
For naturally trained feature extractors, we consistently observe that an optimizer is able to
find perturbations for $x_2$ which significantly reduce the IR-matching distance by an multiple 
orders of magnitude~($10^{-1}$ to $10^{-4}$). 
This shows that for natural feature extractors, we cannot \textit{uniquely} map an IR back to
its original image as there exist multiple preimages that can generate~(approximately) the same IR.
On the other hand, when the robust prior is imposed by an adversarially trained feature-extractor,
the IR-matching distance does not dramatically decrease. Similarly, this implies that the mapping
of $h_F(.;\hat\theta_F)$ is largely one-to-one from the image to the IR. A consequence of this
is it allows \abc to uniquely invert robust IRs to the correct image, enabling data reconstruction.

\subsection{Evaluating IR Leakage Rate from \textsf{SpAB}}
\begin{figure}[t!]
\centering
\begin{subfigure}{0.9\linewidth}
  \includegraphics[width=\linewidth,trim={0em 1.1em 0.0em 0.7em}, clip]{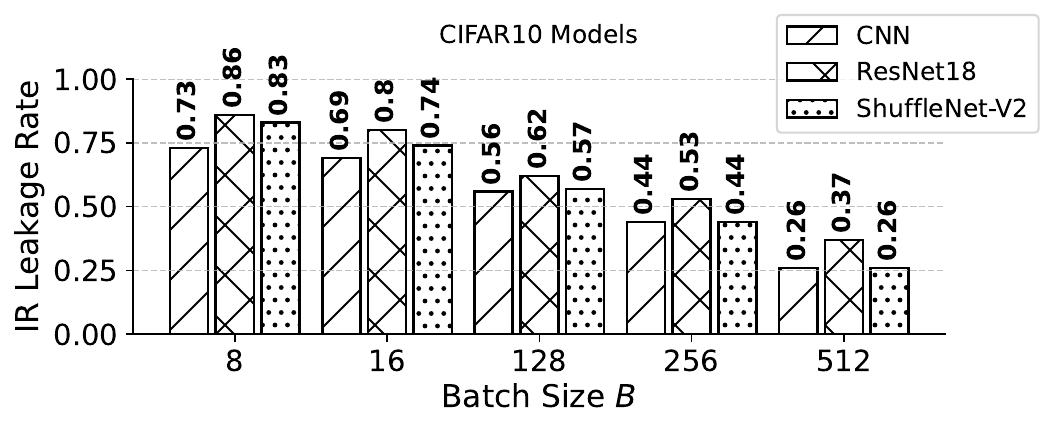}
  \label{fig:leakage-group1}
\end{subfigure}
\begin{subfigure}{0.9\linewidth}
  \includegraphics[width=\linewidth,trim={0em 1.1em 0.0em 0.7em}, clip]{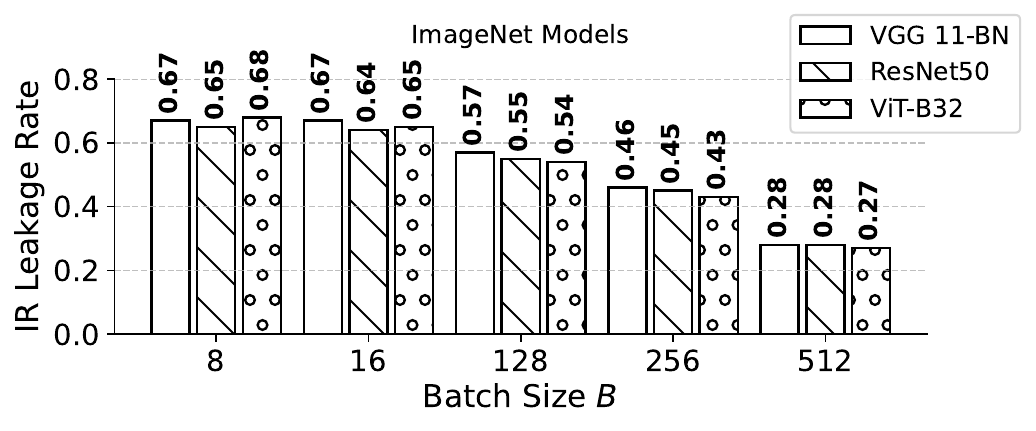}
  \label{fig:leakage-group2}
\end{subfigure}
\vspace{-1em} 
\caption{IR Leakage rate of \abc versus batch size.}
\label{fig:leakage-rate-batch}
\end{figure}
\begin{figure*}[t!]
\centering
\begin{subfigure}{0.25\linewidth}
  \includegraphics[width=\linewidth]{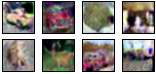}
  \caption{CIFAR10 on CNN}
\end{subfigure}
\hspace{0.5em}
\rulesep
\hspace{0.5em}
\begin{subfigure}{0.25\linewidth}
  \includegraphics[width=\linewidth]{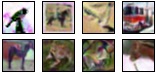}
  \caption{CIFAR10 on ResNet-18}
\end{subfigure}
\hspace{0.5em}
\rulesep
\hspace{0.5em}
\begin{subfigure}{0.25\linewidth}
  \includegraphics[width=\linewidth]{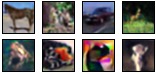}
  \caption{CIFAR10 on ShuffleNet-V2}
\end{subfigure}
\begin{subfigure}{0.25\linewidth}
  \includegraphics[width=\linewidth]{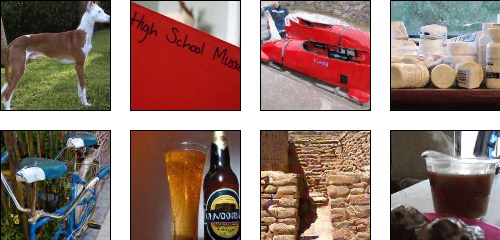}
  \caption{ImageNet on VGG 11-BN}
\end{subfigure}
\hspace{0.5em}
\rulesep
\hspace{0.5em}
\begin{subfigure}{0.25\linewidth}
  \includegraphics[width=\linewidth]{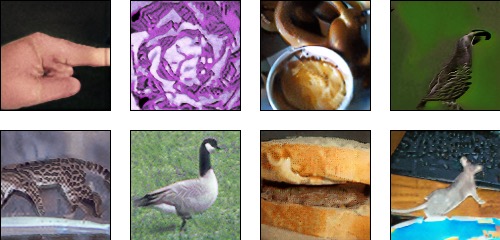}
  \caption{ImageNet on ResNet-50}
\end{subfigure}
\hspace{0.5em}
\rulesep
\hspace{0.5em}
\begin{subfigure}{0.25\linewidth}
  \includegraphics[width=\linewidth]{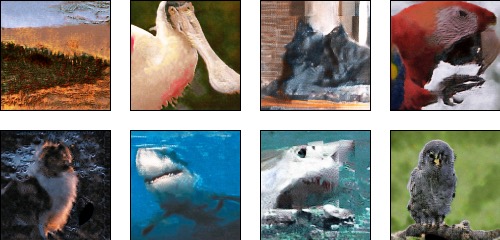}
  \caption{ImageNet on ViT B32}
\end{subfigure}
\caption{Successfully reconstructed samples from randomly sampled batches from the respective
 $D^\textsf{priv}$ datasets.}
\label{fig:recon-viz}
\end{figure*}
We evaluate the IR leakage rate~(\autoref{sec:ir-extraction}) from the first linear
layer of the inserted $\spab(512)$ module.
As a result of \spab-training induced column sparsity, this linear layer exhibits a large IR leakage
rate which implies a larger fraction of the client's batch can be subsequently reconstructed by the server.

\noindent\textbf{IR Leakage Rate with Respect to Batch Size.}
We sample batches of various sizes and generating the gradient update using the cross entropy
loss. We then count the number of distinct IRs which can be recovered to compute the IR leakage rate.
Our results are presented in~\autoref{fig:leakage-rate-batch}.
On CIFAR10 models we observe that the $\spab(512)$ module shows a high leakage rate of 
$27\%-37\%$~(\autoref{tab:spab-conf}) for $B=512$. We also observe a consistent increase
in the IR leakage rate with decreasing batch sizes with a $\sim80\%$ leakage rate for $B=8$. 
On the ImageNet models, the leakage rate for $B=512$ is similar to the leakage rate of the CIFAR10 models.


\subsection{Reconstruction Accuracy}
\input{tables/results-table}
\input{tables/ood-recon}

\begin{figure}[t!]
\centering
\begin{subfigure}{0.75\linewidth}
  \includegraphics[width=\linewidth]{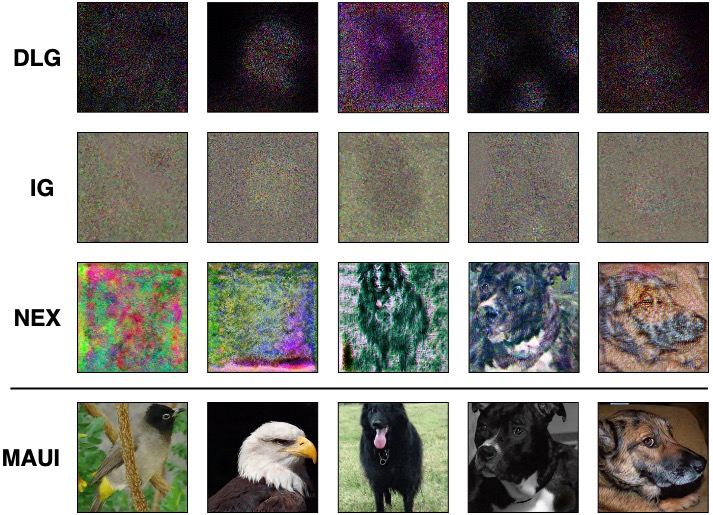}
\end{subfigure}
\caption{Visual comparison of different DRAs on ImageNet samples using IRs from VGG 11-BN for 
batch size $B=1$.}
\label{fig:recon-comp}
\end{figure}

We present reconstruction results of CIFAR10 batches from CNN, ShuffleNet-V2 and ResNet-18 and 
ImageNet batches from VGG11-BN, ResNet-50 and ViT-B32 models. 
We also compare \abc against three prior DRAs.
The results are reported in \autoref{tab:results-cifar} and \autoref{tab:results-imagenet}. 
We select multiple batches from the validation/test splits~($D^\textsf{priv}$) of the same datasets
used for AT and \spab-training. 
These represent reconstructions of in-distribution data samples. 
(Evaluation with out-of-distribution samples is discussed in~\autoref{sec:ood-recon}.)
For each batch, we generate gradients for the $\spab(512)$ block, extract IRs using~\autoref{eqn:ir-extraction}
and perform IR matching using~\autoref{alg:recon}. 
We run reconstructions for multiple batches of different sizes for each configuration. 
For a fair evaluation with baselines, we only consider works which respect our threat 
model and do not use any privacy-leaking primitives. 
This rules out the analytic DRAs~(\autoref{tab:intro-table}) plus Scale-MIA~\cite{shi2025scale}.
Specifically, we consider gradient-based DRAs -- Deep Leakage from Gradients~\cite{zhu2019deep},
Inverting Gradients~(IG)~\cite{geiping2020inverting}, and the IR-based Neuron Exclusivity~(NEX)
~\cite{pan2022exploring} DRA.
For consistency, each attack is evaluated over the same set of batches used to evaluate \abc. 
{\em {\bf In every single case, i.e., all batch sizes, all metrics, all models, and both datasets,
we observe substantial improvements in terms of reconstruction accuracy over prior works.}} 
For ImageNet reconstructions, we consistently achieve PSNR scores $>20$ 
while NEX is limited to $\sim 10$. 
NEX only relies on a naturally trained feature extractor to generate IRs and the DIP for IR-matching
and consequently suffers during reconstruction. 
However, we do observe that NEX is able to generate reconstructions that show similarity in 
textures and large features~(\autoref{fig:recon-comp}).
On the other hand, all considered gradient-based DRAs generate poor reconstructions and artifacts 
with PSNR scores in the range of ~$4-10$. This can be attributed to the lack of gradient information
available to these DRAs as our setting only involves the sharing of \spab gradients. 
We provide a visual comparison of reconstructions from prior DRAs in~\autoref{fig:recon-comp}.

\noindent\textbf{Effect of Batch Size on Reconstruction.}
A significant advantage of \abc is in its agnosticity to batch size. It is important to note that
for \abc, the reconstruction proceeds independently \textit{per IR}, that is, it is not affected
by properties of the batch such as label distribution or batch size. 
This is unlike prior gradient matching works~\cite{zhu2019deep,geiping2020inverting,yin2021see}
which jointly reconstruct the entire batch and suffer when a batch size is large or it contains
multiple samples from the same class~\cite{zhao2024leak}.
For \abc, the effect of batch content is limited to determining the IR-leakage rate. 
This is validated in \autoref{tab:results-cifar}, \autoref{tab:results-imagenet}, 
where the reconstruction does not degrade for larger batch sizes. 

\noindent\textbf{Effect of Model Architecture on Reconstruction.}
The structural complexity of the feature extractor heavily impacts the accuracy of data reconstruction.
DRAs like DLG and IG are ineffective on deep models as the gradient matching optimization task is 
severely underconstrained. These DRAs typically end up generating artifacts when given gradients 
from models like ResNet50, VGG. 
This is evident from the low reconstruction accuracy scores presented in~\autoref{tab:results-imagenet}.
On the other hand, for \abc, we observe that reconstructions on deep architectures are highly accurate.
In fact, VGG-11 shows the highest reconstruction accuracy and is surprisingly able to reconstruct 
minute details like pen-strokes and brick textures~\autoref{fig:recon-viz}.

\subsubsection{Out-of-Distribution Reconstructions}
\label{sec:ood-recon}
\begin{figure}[t!]
\centering
\begin{subfigure}{0.8\linewidth}
  \includegraphics[width=\linewidth]{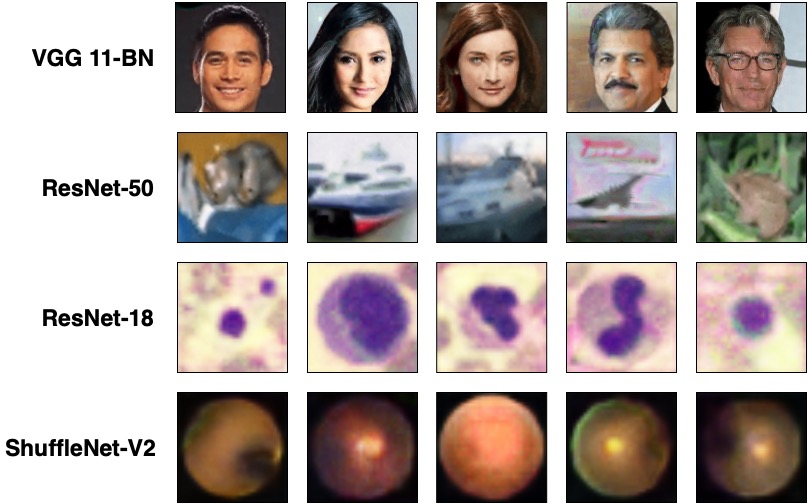}
\end{subfigure}
\caption{OOD Reconstructions on samples from (top-bottom) CelebA, CIFAR10, BloodMNIST and RetinaMNIST.}
\label{fig:ood-recon-viz}
\end{figure}

\abc requires access to a dataset $D^\textsf{pub}$ that is reasonably similar to the client's 
private dataset to learn a transferable image to IR mapping that is largely one to one.
To test the robustness of robust priors, we now consider out-of-distribution reconstructions.
In the evaluations presented earlier, private samples were drawn from the same distribution
as the public dataset used by the server for AT and \spab-training. In this section, we consider
the case when $D^\textsf{priv}$ belongs to a different distribution than $D^\textsf{pub}$. 
For this, we evaluate reconstructions in the following configurations -
$D^{\textsf{pub}}=\text{CIFAR10}$, $D^\textsf{priv}=\{\text{RetinaMNIST, BloodMNIST}\}$ and
$D^{\textsf{pub}}=\text{ImageNet}$, $D^\textsf{priv}=\{\text{CelebA, CIFAR10}\}$.
As we have already established the independence of reconstruction accuracy from the batch size,
we simply evaluate single sample reconstructions~\ie $B=1$ and present the results in \autoref{tab:ood-imagenet}.
The benefits of the robust prior are highlighted perceptually by the accuracy of o.o.d. 
reconstructions(~\autoref{fig:ood-recon-viz}). Despite human faces not being a prominent feature
in the ImageNet dataset, a robust feature extractor is able to encode high level, recognizable
features from CelebA samples into the IRs. 
As a result we see highly accurate reconstructions for all models~(PSNR $>23$, SSIM $>0.5$). 
Similarly, with models trained on CIFAR10, reconstruction of RetinaMNIST samples is achieved with 
high accuracy~(SSIM $>0.8$).
\begin{figure}[t!]
\centering
\begin{subfigure}{0.48\linewidth}
  \includegraphics[width=\linewidth,trim={0em 0em 0em 0em}, clip]{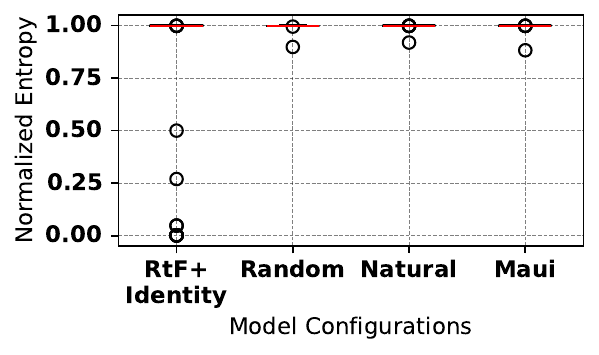}
\end{subfigure}
\begin{subfigure}{0.49\linewidth}
  \centering
  \resizebox{0.99\textwidth}{!}{%
  \begin{tabular}{ccc}
    \hline
    \textbf{\begin{tabular}[c]{@{}c@{}}Model \\ Configuration\end{tabular}} & 
    \textbf{\begin{tabular}[c]{@{}c@{}}Minimum\\ Entropy\end{tabular}} & 
    \textbf{\begin{tabular}[c]{@{}c@{}}$3^{rd}$\\percentile \end{tabular}} \\ \hline
    RtF + Identity & $9.9\times 10^{-4}$ & 0.17 \\ \hline
    Random Model & $8.9\times 10^{-1}$ & 0.99 \\ \hline
    Natural Model & $9.1\times 10^{-1}$ & 0.99 \\ \hline\hline
    \textbf{\abc} & $8.8\times 10^{-1}$ & 0.99 \\ \hline
  \end{tabular}%
  }
  \vspace*{1.5em}
\end{subfigure}
\caption{Boxplot statistics for the normalized entropy value of layer weights in different 
configurations of a CNN.}
\label{fig:stats-anomaly} 
\end{figure}
\begin{table*}[t!]
\centering
\caption{Reconstruction for $(\varepsilon, 10^{-4})$-DP on the CIFAR10 dataset with batch size $B=8$.}
\label{tab:dp-results}
\resizebox{0.9\textwidth}{!}{%
\renewcommand{\arraystretch}{1.1}
\begin{tabular}{ccccccccccccc}
\hline
\multirow{2}{*}{\textbf{$\varepsilon$}} &
\multicolumn{4}{c}{\textbf{CNN $(S_f=100)$}} &
\multicolumn{4}{c}{\textbf{ResNet-18 $(S_f=150)$}} &
\multicolumn{4}{c}{\textbf{ShuffleNet-V2 $(S_f=40)$}} \\
&
PSNR $\uparrow$ & SSIM $\uparrow$ & LPIPS $\downarrow$ & \multicolumn{1}{c|}{Rate $\uparrow$} &
PSNR $\uparrow$ & SSIM $\uparrow$ & LPIPS $\downarrow$ & \multicolumn{1}{c|}{Rate $\uparrow$} &
PSNR $\uparrow$ & SSIM $\uparrow$ & LPIPS $\downarrow$ & Rate $\uparrow$ \\ \hline
$10^6$ & $18.32 \pm 3.36$ & $0.44 \pm 0.14$ & $0.05 \pm 0.03$ & \multicolumn{1}{c|}{$0.58$} & $18.67 \pm 3.35$ & $0.56 \pm 0.12$ & $0.06 \pm 0.06$ & \multicolumn{1}{c|}{$0.71$} & $19.87 \pm 1.98$ & $0.60 \pm 0.11$ & $0.04 \pm 0.02$ & $0.75$ \\ \hline
$10^5$ & $17.21 \pm 2.60$ & $0.42 \pm 0.10$ & $0.06 \pm 0.04$ & \multicolumn{1}{c|}{{\color{blue}$0.58$}} & $18.65 \pm 2.90$ & $0.56 \pm 0.12$ & $0.05 \pm 0.03$ & \multicolumn{1}{c|}{{\color{blue}$0.79$}} & $19.71 \pm 2.17$ & $0.62 \pm 0.11$ & $0.04 \pm 0.02$ & {\color{blue}$0.83$} \\ \hline
$10^4$ & $18.73 \pm 2.04$ & $0.51 \pm 0.14$ & $0.06 \pm 0.03$ & \multicolumn{1}{c|}{$0.42$} & $17.84 \pm 2.41$ & $0.52 \pm 0.11$ & $0.06 \pm 0.04$ & \multicolumn{1}{c|}{$0.67$} & $19.02 \pm 1.52$ & $0.55 \pm 0.13$ & $0.06 \pm 0.03$ & $0.71$ \\ \hline
$10^3$ & $17.00 \pm 0.63$ & $0.39 \pm 0.06$ & $0.08 \pm 0.03$ & \multicolumn{1}{c|}{$0.25$} & $16.54 \pm 2.98$ & $0.49 \pm 0.18$ & $0.09 \pm 0.05$ & \multicolumn{1}{c|}{$0.42$} & $15.76 \pm 2.54$ & $0.42 \pm 0.19$ & $0.10 \pm 0.06$ & $0.50$ \\ \hline
$10^1$ & {\color{gray}$10.83 \pm 1.56$} & {\color{gray}$0.06 \pm 0.06$} & {\color{gray}$0.25 \pm 0.04$} & \multicolumn{1}{c|}{{\color{gray} --}} & {\color{gray} $10.47 \pm 1.53$} & {\color{gray} $0.06 \pm 0.06$} & {\color{gray} $0.25 \pm 0.07$} & \multicolumn{1}{c|}{{\color{gray} --}} & {\color{gray} $10.40 \pm 1.52$} & {\color{gray} $0.06 \pm 0.05$} & {\color{gray} $0.26 \pm 0.06$} & {\color{gray} --} \\ \hline
\end{tabular}%
}
\end{table*}%

\subsection{Evaluation Against an Active Client}
\subsubsection{Normalized Entropy based Anomaly Detection}
\label{sec:anomaly-detection}
We present additional experiments on the normalized-entropy mechanism described in~\autoref{sec:anomaly-detector},
for detecting privacy-leaking primitives. 
To further verify whether the entropy values observed in~\autoref{sec:anomaly-detector} occur in
benign models, we compute the normalized entropies for a CNN in four configurations -- 
with the RtF primitive as $h_C$ and three identity-kernel primitives in each convolution layer, 
with a $\spab(512)$ as $h_C$, 
with randomly initialized weights and a naturally trained model.
We compute the normalized entropy for each layer's weight vector(s) and the boxplot statistics 
over the entire model. The results are presented in~\autoref{fig:stats-anomaly}. 
When compared to other configurations, RtF+Identity is the only configuration that exhibits outliers
with unusually low entropy values~($<0.5$).
3\% of weight vectors~(corresponding to the primitives) have an entropy $<0.17$ which correspond
to the weights of the layers with the privacy-leaking primitives.
This low entropy is principally because of the presence of nearly identical parameter values
(within a tolerance of $10^{-6}$), which does not occur during natural training, random initialization
or during \spab-training.
Thus, \abc is undetectable to the anomaly-detection mechanism and is thus stealthier than prior works.

\subsubsection{Evaluation Against Local Differential Privacy}
We consider the standard $(\varepsilon, \delta)-$DP mechanism. The client uses this to perturb 
the gradient update before sending it to the server. LDP performs two operations on the gradient
to ensure privacy of the underlying data --- {\em (i)} clipping, which scales the gradient by its 
$L_2$ norm to a user specified clipping threshold~($S_f$) and {\em (ii)} noising, which adds 
gaussian noise of a fixed variance to the update. The variance of noise is determined using a privacy 
budget~($\varepsilon$). We perform \abc on gradients perturbed under different
privacy budgets using a single-shot DP mechanism.
A sample is said to be successfully reconstructed if SSIM $>3.0$. The reconstruction rate is 
computed as the fraction of the batch reconstructed successfully.

\noindent\textbf{\spab-training with Incorrect Labels.}
We observe that the magnitudes of the weight and bias gradient norms shows extreme variance 
between batch samples that are classified correctly at the client versus those that are misclassified. 
Choosing a small $S_f$ leads to significant clipping of gradients with large norms while for a 
large $S_f$, significant noise destroys the information content. 
This significantly degrades the reconstruction rate for the server. To overcome this issue, we employ
a trick during \spab-training by mislabeling the training data.
This causes most samples to be misclassified at the client side resulting in large and uniform
gradient norms.
We present our results in~\autoref{tab:dp-results}. Note that reconstruction accuracy is only computed
over samples that are successfully reconstructed.
We see that \abc is able to achieve successful reconstructions for large $\varepsilon$ values.
We also observe that moderate DP actually increases the reconstruction rate~(highlighted in {\color{blue} blue}).
However, strong DP reduces the effectiveness of the attack~({\color{gray} Gray} values are 
unsuccessful reconstruction metrics for $\varepsilon=10$). 
However, we also note that with small privacy budgets, the shared gradients are so inaccurate 
that the utility of model training itself becomes unacceptably low. 

%% file: tables/results-table.tex
\begin{table*}[t!]
\centering
\caption{Comparing reconstruction accuracies on CIFAR-10 dataset with different batch sizes $B$.}
\label{tab:results-cifar}
\resizebox{0.9\textwidth}{!}{%
\renewcommand{\arraystretch}{1.1}
\begin{tabular}{ccccc|ccc|ccc}
\hline
&
&
\multicolumn{3}{c|}{\textbf{CNN}} & \multicolumn{3}{c|}{\textbf{ResNet-18}} & \multicolumn{3}{c}{\textbf{ShuffletNet-V2}} \\
\multirow{-2}{*}{\textbf{Attack}} &
\multirow{-2}{*}{\textbf{$B$}} &
PSNR $\uparrow$ & SSIM $\uparrow$ & LPIPS $\downarrow$ &
PSNR $\uparrow$ & SSIM $\uparrow$ & LPIPS $\downarrow$ &
PSNR $\uparrow$ & SSIM $\uparrow$ & LPIPS $\downarrow$ \\ \hline
&
8 &
$6.88 \pm 0.43$ & $0.01 \pm 0.00$ & $0.29 \pm 0.02$ & $6.08 \pm 0.75$ & $0.00 \pm 0.00$ &
$0.34 \pm 0.02$ & $6.28 \pm 0.10$ & $0.00 \pm 0.00$ & $0.33 \pm 0.01$ \\
&
16 &
$6.13 \pm 0.77$ & $0.01 \pm 0.00$ & $0.33 \pm 0.03$ & $6.49 \pm 0.94$ & $0.01 \pm 0.00$ &
$0.35 \pm 0.01$ & $6.29 \pm 0.30$ & $0.00 \pm 0.00$ & $0.35 \pm 0.02$ \\
\multirow{-3}{*}{DLG~\cite{zhu2019deep}} &
32 &
$5.55 \pm 0.44$ & $0.00 \pm 0.00$ & $0.32 \pm 0.01$ & $5.87 \pm 0.59$ & $0.01 \pm 0.00$ &
$0.36 \pm 0.03$ & $6.06 \pm 0.02$ & $0.01 \pm 0.00$ & $0.36 \pm 0.02$ \\ \hline
&
8 &
$12.54 \pm 1.00$ & $0.02  \pm 0.01$ & $0.21 \pm 0.01$ & $10.69 \pm 0.86$ & $0.00 \pm 0.00$ &
$0.27  \pm 0.04$ & $10.46 \pm 0.46$ & $0.02 \pm 0.02$ & $0.46  \pm 0.02$ \\ 
&
16 &
$12.35 \pm 0.56$ & $0.01  \pm 0.00$ & $0.24 \pm 0.02$ & $10.41 \pm 0.29$ & $0.01 \pm 0.00$ &
$0.34  \pm 0.03$ & $10.64 \pm 0.64$ & $0.01 \pm 0.02$ &  $0.51 \pm 0.04$ \\
\multirow{-3}{*}{IG~\cite{geiping2020inverting}} &
32 &
$11.57 \pm 0.36$ & $0.00  \pm 0.00$ & $0.27 \pm 0.02$ & $10.52 \pm 0.46$ & $0.01 \pm 0.01$ &
$0.40  \pm 0.01$ & $10.17 \pm 0.32$ & $0.00 \pm 0.01$ & $0.51  \pm 0.01$ \\ \hline &
8 &
$10.55 \pm 1.69$ & $0.09  \pm 0.09$ & $0.21 \pm 0.09$ & $13.42 \pm 1.53$ & $0.29 \pm 0.14$ &
$0.12  \pm 0.05$ & $11.41 \pm 2.29$ & $0.14 \pm 0.14$ & $0.19  \pm 0.07$ \\
&
16 &
$12.30 \pm 2.44$ & $0.17  \pm 0.14$ & $0.14 \pm 0.09$ & $13.92 \pm 1.77$ & $0.27 \pm 0.12$ &
$0.12  \pm 0.05$ & $12.26 \pm 2.42$ & $0.15 \pm 0.12$ &  $0.19 \pm 0.10$ \\
\multirow{-3}{*}{NEX~\cite{pan2022exploring}} &
32 &
$11.94 \pm 3.41$ & $0.15  \pm 0.18$ & $0.20 \pm 0.11$ & $14.37 \pm 2.81$ & $0.30 \pm 0.15$ &
$0.11  \pm 0.06$ & $12.38 \pm 2.19$ & $0.18 \pm 0.13$ & $0.19  \pm  0.09$ \\ \hline\hline
\rowcolor[HTML]{EFEFEF} 
\cellcolor[HTML]{EFEFEF} &
8 &
$17.32 \pm 4.25$ & $0.437 \pm 0.18$ & $0.05 \pm 0.04$ & $22.01 \pm 3.68$ & $0.71 \pm 0.12$ &
$0.03  \pm 0.03$ & $20.66 \pm 3.43$ & $0.66 \pm 0.22$ &  $0.04 \pm 0.04$ \\
\rowcolor[HTML]{EFEFEF} 
\cellcolor[HTML]{EFEFEF} &
16 &
$17.59 \pm 3,15$ & $0.47  \pm 0.18$ & $0.06 \pm 0.05$ & $21.28 \pm 2.90$ & $0.70 \pm 0.13$ &
$0.04  \pm 0.02$ & $20.92 \pm 2.50$ & $0.66 \pm 0.13$ & $0.03  \pm 0.02$ \\
\rowcolor[HTML]{EFEFEF} 
\multirow{-3}{*}{\cellcolor[HTML]{EFEFEF}\textbf{\abc}} &
32 &
$17.40 \pm 4.39$ & $0.45  \pm 0.20$ & $0.08 \pm 0.06$ & $21.73 \pm 3.81$ & $0.70 \pm 0.19$ &
$0.04  \pm 0.04$ & $21.72 \pm 3.29$ & $0.72 \pm 0.10$ & $0.03  \pm 0.02$ \\ \hline
\end{tabular}%
}
\end{table*}%
\begin{table*}[t!]
\centering
\caption{Comparing reconstruction accuracies on ImageNet dataset with different batch sizes $B$.}
\label{tab:results-imagenet}
\resizebox{0.9\textwidth}{!}{%
\renewcommand{\arraystretch}{1.1}
\begin{tabular}{ccccc|ccc|ccc}
\hline
&
&
\multicolumn{3}{c|}{\textbf{VGG 11-BN}} & \multicolumn{3}{c|}{\textbf{ResNet-50}} & \multicolumn{3}{c}{\textbf{ViT B-32}} \\
\multirow{-2}{*}{\textbf{Attack}} &
\multirow{-2}{*}{\textbf{$B$}} &
PSNR $\uparrow$ & SSIM $\uparrow$ & LPIPS $\downarrow$ &
PSNR $\uparrow$ & SSIM $\uparrow$ & LPIPS $\downarrow$ &
PSNR $\uparrow$ & SSIM $\uparrow$ & LPIPS $\downarrow$ \\ \hline
&
1 &
$10.95 \pm 1.51$ & $0.08 \pm 0.05$ & $0.96 \pm 0.13$ & $5.70 \pm 1.77$ & $0.01 \pm 0.01$ &
$1.30 \pm 0.12$ & $5.97 \pm 1.21$ & $0.00 \pm 0.00$ & $1.34 \pm 0.06$ \\
&
4 &
$7.04 \pm 0.06$ & $0.05 \pm 0.01$ & $0.86 \pm 0.08$ & $4.51 \pm 0.12$ & $0.00 \pm 0.00$ &
$1.34 \pm 0.04$ & $5.15 \pm 0.78$ & $0.00 \pm 0.00$ & $1.35 \pm 0.00$ \\
\multirow{-3}{*}{DLG~\cite{zhu2019deep}} &
8 &
$7.16 \pm 0.68$ & $0.03 \pm 0.03$ & $0.99 \pm 0.28$ & $6.07 \pm 0.82$ & $0.00 \pm 0.00$ & 
$1.32 \pm 0.03$ & $5.24 \pm 0.77$ & $0.00 \pm 0.00$ &  $1.34 \pm 0.03$ \\ \hline
&
1 &
$10.85 \pm 2.45$ & $0.00 \pm 0.00$ & $1.11 \pm 0.15$ & $10.14 \pm 2.45$ & $0.01 \pm 0.01$ &
$0.90 \pm 0.08$ & $10.18 \pm 2.05$ & $0.00 \pm 0.00$ & $1.10 \pm 0.09$ \\
&
4 &
$10.73 \pm 0.46$ & $0.00 \pm 0.00$ & $1.08 \pm 0.02$ & $10.02 \pm 0.56$ &  $0.02 \pm 0.03$ & 
$0.84 \pm 0.02$ & $10.25 \pm 0.37$ & $0.00 \pm 0.00$ & $0.99 \pm 0.02$ \\
\multirow{-3}{*}{IG~\cite{geiping2020inverting}} &
8 &
$11.23 \pm 0.33$ & $0.00 \pm 0.00$ & $1.08 \pm 0.02$ & $10.17 \pm 0.16$ & $0.03 \pm 0.03$ &
$0.87 \pm 0.08$ & $10.79 \pm 0.43$ & $0.00 \pm 0.00$ & $0.98 \pm 0.02$ \\ \hline
&
1 &
$10.56\pm1.66$ & $0.06\pm0.02$ & $0.82\pm0.25$ & $11.15\pm1.76$ & $0.10\pm0.02$ &
$0.61\pm0.08$ & $9.47\pm1.52$ & $0.05\pm0.02$  &  $0.76\pm0.09$ \\
&
4 &
$11.04\pm0.64$ & $0.07\pm0.01$ & $0.76\pm0.13$ & $11.42\pm1.34$ & $0.15\pm0.07$ &
$0.68\pm0.04$ & $10.22\pm1.19$ & $0.06\pm0.03$ & $0.89\pm0.12$ \\
\multirow{-3}{*}{NEX~\cite{pan2022exploring}} &
8 &
$11.59\pm1.58$ & $0.10\pm0.03$ & $0.67\pm0.07$ & $11.21\pm1.60$ & $0.12\pm0.07$ &
$0.76\pm0.20$ & $10.31\pm1.02$ & $0.07\pm0.04$ & $0.84\pm0.14$ \\ \hline\hline
\rowcolor[HTML]{EFEFEF} 
\cellcolor[HTML]{EFEFEF} &
1 &
$24.17\pm3.40$ & $0.56\pm0.13$ & $0.11\pm0.03$ & $20.72\pm2.24$ & $0.41\pm0.12$ &
$0.26\pm0.06$ & $18.26\pm2.45$ & $0.31\pm0.14$ & $0.37\pm0.12$ \\
\rowcolor[HTML]{EFEFEF} 
\cellcolor[HTML]{EFEFEF} &
4 &
$25.42\pm3.52$ & $0.65\pm0.18$ & $0.11\pm0.04$ & $18.64\pm6.18$ & $0.41\pm0.23$ &
$0.35\pm0.24$ & $17.32\pm5.07$ & $0.36\pm0.19$ & $0.40\pm0.10$ \\
\rowcolor[HTML]{EFEFEF} 
\multirow{-3}{*}{\cellcolor[HTML]{EFEFEF}\textbf{\abc}} &
8 &
$24.00\pm4.84$ & $0.62\pm0.19$ & $0.11\pm0.05$ & $19.29\pm3.64$ & $0.44\pm0.20$ &
$0.25\pm0.10$ & $18.75\pm4.02$ & $0.39\pm0.20$ & $0.32\pm0.11$ \\ \hline
\end{tabular}%
}
\end{table*}%

%% file: tables/ood-recon.tex
\begin{table*}[t!]
\caption{\abc's Reconstruction Accuracy for Out-of-Distribution samples. 
$D^\textsf{pub}$ is CIFAR10~(top), ImageNet~(bottom).
}
\centering
\resizebox{0.9\linewidth}{!}{%
\begin{tabular}{cccccccccc}
\hline
\multirow{2}{*}{\textbf{$D^\textsf{priv}$}} & \multicolumn{3}{c}{\textbf{CNN}} & \multicolumn{3}{c}{\textbf{ResNet-18}} & \multicolumn{3}{c}{\textbf{ShuffleNet-V2}} \\
            & PSNR~$\uparrow$ & SSIM~$\uparrow$ & \multicolumn{1}{c|}{LPIPS~$\downarrow$} & PSNR~$\uparrow$ & SSIM~$\uparrow$ & \multicolumn{1}{c|}{LPIPS~$\downarrow$} & PSNR~$\uparrow$ & SSIM~$\uparrow$ & LPIPS~$\downarrow$ \\ \hline
BloodMNIST  & $24.94\pm0.79$  & $0.80\pm0.05$ & \multicolumn{1}{c|}{$0.02\pm0.01$}   & $25.55\pm1.32$  & $0.81\pm0.08$  & \multicolumn{1}{c|}{$0.02\pm0.02$}   & $24.94\pm0.79$ & $0.80\pm0.05$  & $0.02\pm0.01$   \\ \hline
RetinaMNIST & $27.89\pm1.55$  & $0.84\pm0.05$ & \multicolumn{1}{c|}{$0.01\pm0.00$}   & $30.17\pm1.67$  & $0.88\pm0.04$  & \multicolumn{1}{c|}{$0.01\pm0.01$}   & $30.02\pm0.62$ & $0.90\pm0.01$ & $0.01\pm0.01$   \\ \hline
\end{tabular}%
}
\vspace {1 em}

\resizebox{0.9\linewidth}{!}{%
\begin{tabular}{cccccccccc}
\hline
\multirow{2}{*}{\textbf{$D^\textsf{priv}$}} & \multicolumn{3}{c}{\textbf{VGG 11-BN}} & \multicolumn{3}{c}{\textbf{ResNet-50}} & \multicolumn{3}{c}{\textbf{ViT B-32}} \\
        & PSNR~$\uparrow$ & SSIM~$\uparrow$ & \multicolumn{1}{c|}{LPIPS~$\downarrow$} & PSNR~$\uparrow$ & SSIM~$\uparrow$ & \multicolumn{1}{c|}{LPIPS~$\downarrow$} & PSNR~$\uparrow$ & SSIM~$\uparrow$ & LPIPS~$\downarrow$ \\ \hline
CelebA  & $28.66\pm1.46$  & $0.78\pm0.06$  & \multicolumn{1}{c|}{$0.08\pm0.03$}   & $23.83\pm1.57$  & $0.65\pm0.08$  & \multicolumn{1}{c|}{$0.17\pm0.03$}   & $20.44\pm1.27$  & $0.52\pm0.05$  & $0.28\pm0.06$   \\ \hline
CIFAR10 & $37.98\pm1.56$  & $0.95\pm0.02$  & \multicolumn{1}{c|}{$0.09\pm0.03$}   & $26.44\pm5.82$  & $0.76\pm0.15$  & \multicolumn{1}{c|}{$0.28\pm0.21$}   & $24.46\pm4.72$  & $0.73\pm0.13$  & $0.29\pm0.06$   \\ \hline
\end{tabular}%
}
\label{tab:ood-imagenet}
\end{table*}

%% file: sections/8-discussion.tex
\section{Discussion and Future Work}
\label{sec:discussion}
\noindent\textbf{Stealthiness of the \spab module.}
The structure of the \spab module is identical to the RtF~\cite{fowl2022robbing} primitive. However,
unlike RtF, \abc maintains stealthiness through three mechanisms. First, the integration of the \spab module
maintains compatibility with established model architecture paradigms. The arrangement of feature extractors
followed by MLPs is a prevalent and well-accepted design pattern in models such as AlexNet and VGG.
Moreover, the practice of model finetuning often involves replacement of the classification head with 
MLPs. Second, this modification takes place \textit{prior} to federated finetuning following which the 
model architecture remains unchanged. Also note that models like AlexNet and VGG already contain large
MLPs in their classification heads and therefore require no modification. Finally, the process of 
\spab-training learns weights that do not possess a handcrafted nature and appear organically learned.
They do not trigger the entropy based anomaly detector presented in~\autoref{sec:anomaly-detector}.

\noindent\textbf{Effectiveness of Robust IR-matching.}
IR-matching is a first order optimization problem and is generally more stable than gradient matching which
requires computing a gradient of gradients~($\nabla_x\nabla_\theta L(h(x;\theta), c)$). As a result,
it generally leads to more accurate reconstructions as shown by~\cite{pan2022exploring}.
Robust IR-matching significantly improves this reconstruction performance. Interestingly, we observe
that the feature extractor does not need to achieve state-of-the-art robustness to enable
highly accurate reconstructions. In our experiments, models reached moderate robust accuracy levels which
were sufficient for the reconstructions presented in this work~(\autoref{sec:evaluation}).

\noindent\textbf{Limitations and Future Directions.}
\abc suffers two key limitations which lead to interesting directions for future work. First, AT requires access to a considerably large dataset which is representative of the client's private dataset. An improvement to \abc would be to identify strong priors that can be enforced on IRs with a fraction of the data. In the same vein as~\cite{zhao2024leak}, methods which leverage already leaked data to increase leakage rate and reconstruction accuracy could also be explored. Another line of future work involves improving the sparsity objective to  enhance the utility of \spab-training.

%% file: sections/9-conclusion.tex
\section{Conclusion}
We present \abc, an IR based DRA tailored to FTL. It is based on two key insights -- first, it imposes a robust prior on the IR of a sample generated by the feature extractor. Second, it uses a custom sparsity objective to learn parameters for the classification head that allows for large scale extraction of these IRs from its gradient. The action of the robust prior results in highly accurate reconstructions which are significantly better than prior DRAs even on deep models like ResNet50, ViT B32 and its use of IRs makes \abc largely agnostic to the  batch size. Finally, \abc does not use any privacy leaking primitives and is thus stealthier than prior works.

%% file: appendix/appendix.tex
\section{Reconstruction Accuracy Metrics}
\label{app:reconstruction-metrics}
We pick commonly used image similarity metrics to quantify the similarity of the reconstructions with the actual images
\begin{enumerate}[leftmargin=*,itemsep=0pt, parsep=0pt, nosep]
  \item \textit{Peak Signal-to-Noise Ratio~(PSNR) --} PSNR is an extensively used metric which quantifies reconstruction similarity on a logarithmic scale. Formally, it is computed as $\text{PSNR}(x^*, x_g) = -10\times \log_{10}\text{MSE}(x^*, x_g)$. A higher PSNR score generally indicates better reconstruction.
  \item \textit{Structural Similarity~(SSIM) --} SSIM uses the luminance, contrast and structural variance between the reconstructed image and ground truth to capture their similarity. This makes it more aligned with human perception than SNR which focuses on pixelwise differences. SSIM lies in $[-1,1]$ with larger values signifying higher similarity.
  \item \textit{Learned Perceptual Image Patch Similarity~(LPIPS) --} Zhang~\etal~\cite{zhang2018perceptual} proposed visual representations from deep networks as approximators of human perception and used them to compare the differences between two image samples. LPIPS is a loss function with lower values signifying better reconstruction accuracy. We compute the LPIPS loss using the AlexNet trunk for our evaluations.
\end{enumerate}